\documentclass[11pt,a4paper]{article}
\usepackage{amssymb}

\newcounter{bla}

\usepackage{lscape} 
\usepackage{multirow}
\usepackage{braket}
\usepackage{amssymb}
\usepackage{dutchcal}
\usepackage{ifsym}
\usepackage[T1]{fontenc} 
\usepackage{bm}
\usepackage{bbm}
\usepackage{slashed}
\usepackage{longtable}
\usepackage{pifont}

\usepackage{graphicx}%
\usepackage{caption}
\usepackage{subcaption}
\usepackage{tikz}
\usetikzlibrary{decorations.pathmorphing}
\usetikzlibrary{decorations.markings}
\usetikzlibrary{calc}
\usetikzlibrary{math}
\usetikzlibrary{fpu}
\usepackage{standalone}
\usepackage{pgffor}
\usepackage{booktabs} 
\usepackage{amsmath}  

\usepackage{hhline}

\usepackage{parskip}

\usepackage[sort&compress,numbers]{natbib}

\usepackage{lmodern}
\usepackage{mmacells}
\usepackage[most]{tcolorbox}
\usepackage{xcolor}
\definecolor{myGreen}{rgb}{0.2,0.72,0.2}
\definecolor{grey}{rgb}{0.5,0.5,0.5}
\definecolor{myWhite}{rgb}{0.98,0.98,0.98}
\definecolor{myGray}{rgb}{0.7,0.7,0.75}
\definecolor{myGold}{rgb}{0.8,0.64,0.24}

\definecolor{mygreen}{rgb}{0,0.6,0}
\definecolor{mygray}{rgb}{0.5,0.5,0.5}
\definecolor{mymauve}{rgb}{0.58,0,0.82}
\newcommand{\e}{\varepsilon}
\newcommand{\ta}{\left(}
\newcommand{\qa}{\left[}

\newcommand{\tc}{\right)}
\newcommand{\qc}{\right]}

\renewcommand{\[}{\begin{equation}}
\renewcommand{\]}{\end{equation}}

\newcommand{\nn}{\nonumber}

\mmaDefineMathReplacement[≤]{<=}{\leq}
\mmaDefineMathReplacement[≥]{>=}{\geq}
\mmaDefineMathReplacement[≠]{!=}{\neq}
\mmaDefineMathReplacement[→]{->}{\to}[2]
\mmaDefineMathReplacement[⧴]{:>}{:\hspace{-.2em}\to}[2]
\mmaDefineMathReplacement{∉}{\notin}
\mmaDefineMathReplacement{∞}{\infty}
\mmaDefineMathReplacement{��}{\mathbbm{d}}

\mmaSet{
  morefv={gobble=2},
  linklocaluri=mma/symbol/definition:#1,
  morecellgraphics={yoffset=1.9ex}
}

\makeatletter
\newcommand{\oset}[3][0ex]{%
  \mathrel{\mathop{#3}\limits^{
    \vbox to#1{\kern-2\ex@
    \hbox{$\scriptstyle#2$}\vss}}}}
\makeatother

\newcommand{\whiteline}{\vskip 0.5cm}

\usepackage{geometry}
\begin{document}
\allowdisplaybreaks

\vspace{-2.0cm}
\begin{flushright}
  DESY-24-060\\IPARCOS-UCM-24-025
\end{flushright}
\vspace{.1cm}

\begin{center} 

  {\Large \bf  {Numerical implementation of evolution equations for twist-3 collinear PDFs}}
  \vspace{.7cm}

  Simone~Rodini$^a$\footnote{\href{mailto:simone.rodini@desy.de}{simone.rodini@desy.de}}
  Lorenzo~Rossi$^{b,c}$\footnote{\href{mailto:lorenzo.rossi@pv.infn.it}{lorenzo.rossi@pv.infn.it}}
  Alexey~Vladimirov$^d$\footnote{\href{mailto:alexeyvl@ucm.es}{alexeyvl@ucm.es}}

\setcounter{footnote}{0}

\vspace{.3cm}
$^a${\it Deutsches Elektronen-Synchrotron DESY, Notkestr. 85, 22607 Hamburg, Germany}\\
$^b${\it Dipartimento di Fisica, Universit\`a di Pavia, via Bassi 6, Pavia, I-27100, Italy}\\
$^c${\it INFN Sezione di Pavia, via Bassi 6, via Bassi 6, Pavia, I-27100, Italy}\\
$^d${\it Departamento de F\'{i}sica Te\'{o}rica \& IPARCOS, Universidad Complutense de Madrid, E-28040 Madrid,
Spain}\\

\end{center}   

\begin{center}
  {\bf \large Abstract}\\
\end{center}
Twist-3 collinear parton distribution functions (PDFs) are matrix elements of quark-gluon-quark or three-gluons light-cone operators. They depend on three momentum fraction variables, which are restricted to a hexagon region, and the evolution kernels are defined via two-dimensional convolution in these variables. We present the numerical realisation of the twist-3 evolution equations at leading order in the strong coupling for all kinds of twist-3 PDF (quark, gluon, chiral-even/odd, etc). We provide two independent codes (in C and Fortran) that have been extensively cross-checked, and are ready-to-use. We supplement the paper with a review of known properties of twist-3 PDFs.\\
\whiteline
\textbf{keywords}: Hadron structure; Parton distribution functions; evolution equations; twist-3;

\newpage

\tableofcontents

\section{Program summary}

\begin{small}
\noindent
{\bf Programs Titles:}   Honeycomb ({\tt C}) / Snowflake ({\tt FORTRAN})                                     \\
{\bf Code repositories:} \cite{honeycomb,snowflake} \\
{\bf Licensing provisions:} GNU GPL v2.0 ({\tt C}) / MIT ({\tt FORTRAN})  \\
{\bf Nature of the problem:} 
The aim is to develop a code that allows for the evaluation of the evolution equations for the twist-3 parton distributions at LO in the strong coupling with a variable number of flavors. Standard techniques that have been developed for the twist-2 case cannot be straightforwardly applied because of the increased dimensionality of the problem. The balance between performance, memory requirements and accuracy becomes even more important and difficult to achieve. \\
{\bf Solution method:} 1) The choice of coordinates. We parametrize the plane defined by momentum conservation $x_1+x_2+x_3=0$ using radial-angular coordinates in the infinite norm (rather than the Euclidean norm), since it respects the symmetries of the problem. 2) We discretize the radial and angular coordinates. Different grids are provided, the radial grids especially are chosen to have a high density near the origin. 3) The parton distributions are linearly interpolated in grid space. 4) The coupled integro-differential evolution equations are transformed into a system of coupled ordinary differential equations and are solved using the 4th-order Runge-Kutta method.\\
The solution comes in two independent implementations written in the {\tt C} and {\tt FORTRAN} programming languages, which have been cross-checked to ensure the reliability of the final results.\\
\end{small}

\section{Introduction}
\label{sec_introduction}

Parton distribution functions (PDFs) are the cornerstone of modern hadron physics. They parameterize hadron matrix elements of light-cone QCD operators and serve as universal blocks for describing high-energy cross-sections. PDFs depend on dimensionless momentum fraction(s) $x$, which parameterize parton motion, and the renormalization scale $\mu$, which is often associated with the energy scale of the proton. The dependence on $x$ is the subject of modeling and fits, whereas the scale dependence is dictated by the so-called evolution equations \cite{Gribov:1972ri, Altarelli:1977zs}. The derivation and observation of evolution for PDFs is one of the great successes of QCD. The incorporation of evolution into the analysis of data is of ultimate importance. 

PDFs arise in the factorization theorems that are organized as a power expansion over a hard scale. The leading term is governed by so-called twist-2 PDFs, which are known very precisely nowadays. The following power correction is described by the twist-3 PDFs, whose values are practically unknown nowadays. This work is dedicated to twist-3 PDFs, and one of the first steps towards their determination from the data. Twist-3 PDFs parametrize the three-point correlators inside hadrons and are truly quantum objects. For this reason, their exploration is interesting and will grant an understanding of the quantum effects inside the hadrons. Apart from subleading power correction, twist-3 PDFs also appear in the description of many observables already at leading power. The most important examples are structure function $g_2$ \cite{Anselmino:1994gn, Kodaira:1994ge}, single-spin asymmetries \cite{Ji:2006ub, Ji:2006vf}, polarized transverse momentum distributions \cite{Scimemi:2018mmi, Rodini:2022wki,delCastillo:2023rng}, quasi-distributions \cite{Braun:2021gvv, Braun:2021aon} and many others.

The theory of twist-3 PDFs was developed long ago \cite{Bukhvostov:1985rn, Balitsky:1987bk}, but it is almost never applied in practice. The main reason is the absence of data with sufficient quality. With time, the amount of collected data is growing, and presently one may plan a global analysis of multiple data sources, in this way, overcoming the uncertainty issue of individual measurements. It should also be noted that future experiments (such as Electron-Ion collider \cite{AbdulKhalek:2021gbh} or LHCspin \cite{Aidala:2019pit}) are aimed to measure many twist-3-related effects with increased precision. A global analysis is conceptually feasible since many observables are described in terms of a relatively small set of twist-3 PDFs. A necessary element of such analysis is a fast and accurate evolution code, which allows to consistently transform PDFs from one energy scale to another. To our knowledge, only one publicly available code is presented in ref.\cite{Pirnay:2013fra}. However, it was the first exploratory attempt, and it suffered from low accuracy, long computation times and poor integrability in larger codes, all of which are essential aspects to be taken into account for the goal of a global extraction.  

In this work, we present the evolution code for twist-3 PDFs at the leading order (which is the maximum available order at the time of the work). It can evolve all PDFs of twist-3 (there are four types of quark PDFs and two independent types of gluon PDFs), taking into account mixing between them. The evolution is implemented on a 2D grid in the space of momentum fractions. The grid is chosen such that it supports the natural symmetries of PDFs and is denser in the vicinity of points and lines of rapid change. The precision of the computation is variable and tuned by a user for optimal precision/efficiency ratio. We present two implementations of the same algorithm in C and Fortran. They were done independently and cross-checked.

The paper is organized as follows. In sec.\ref{sec_definitions}, we review the main elements of the twist-3 PDFs, namely, their definition, parameterization, symmetries, and evolution equations. We provide a comprehensive collection of relations between notations and parameterization used in different works. In sec.\ref{sec_discretizaitons}, we present a definition of the grid, as it is the main element of the program. In sec.\ref{sec_results}, we demonstrate the output of the program by providing few examples. The codes details are presented in sec.\ref{sec_code_interface}.

\section{Twist-three parton distribution functions}
\label{sec_definitions}

\begin{figure}[!ht]
\begin{center}
\begin{tikzpicture}
\tikzmath{\s = 1.5;}

\coordinate (e1) at (0.866025*\s, -0.5*\s);
\coordinate (e2) at (0.*\s, 1.*\s);
\coordinate (e3) at (-0.866025*\s, -0.5*\s);

\draw[-latex,thick,red] (0,0) -- ($3*(e1)$);
\draw[-latex,thick,red] (0,0) -- ($2.6*(e2)$);
\draw[-latex,thick,red] (0,0) -- ($3*(e3)$);

\node[right] at ($3*(e1)$) {$x_1$};
\node[right] at ($2.6*(e2)$) {$x_2$};
\node[left] at ($3*(e3)$) {$x_3$};

\node[right,draw=black!50] at ($1.1*(e1)-1.4*(e2)-0*(e3)$) {$x_1+x_2+x_3=0$};

\draw[] ($-1*(e1)+1*(e2)+0*(e3)$) -- 
		($-1*(e1)+0*(e2)+1*(e3)$) --
		($0*(e1)-1*(e2)+1*(e3)$) --
		($1*(e1)-1*(e2)-0*(e3)$) --
		($1*(e1)+0*(e2)-1*(e3)$) --
		($0*(e1)+1*(e2)-1*(e3)$) --
		($-1*(e1)+1*(e2)+0*(e3)$);

\draw[] (0,0)--($-1*(e1)+1*(e2)+0*(e3)$);
\draw[] (0,0)--($-1*(e1)+0*(e2)+1*(e3)$);
\draw[] (0,0)--($0*(e1)-1*(e2)+1*(e3)$);
\draw[] (0,0)--($1*(e1)-1*(e2)-0*(e3)$);
\draw[] (0,0)--($1*(e1)+0*(e2)-1*(e3)$);
\draw[] (0,0)--($0*(e1)+1*(e2)-1*(e3)$);

\draw[white,fill=white] ($0.6*(e1)-0.3*(e2)-0.3*(e3)$) circle (5pt);
\node[blue!80!green,fill=white,opacity=0.,text opacity=1] at ($0.6*(e1)-0.3*(e2)-0.3*(e3)$) {$\mathbf{(+--)}$};
\node[blue!80!green,fill=white,opacity=0.,text opacity=1] at ($0.3*(e1)-0.6*(e2)+0.3*(e3)$) {$\mathbf{(+-+)}$};
\draw[white,fill=white] ($-0.3*(e1)-0.3*(e2)+0.6*(e3)$) circle (5pt);
\node[blue!80!green,fill=white,opacity=0.,text opacity=1] at ($-0.3*(e1)-0.3*(e2)+0.6*(e3)$) {$\mathbf{(--+)}$};
\node[blue!80!green,fill=white,opacity=0.,text opacity=1] at ($-0.6*(e1)+0.3*(e2)+0.3*(e3)$) {$\mathbf{(-++)}$};
\draw[white,fill=white] ($-0.3*(e1)+0.6*(e2)-0.3*(e3)$) circle (5pt);
\node[blue!80!green,fill=white,opacity=0.,text opacity=1] at ($-0.3*(e1)+0.6*(e2)-0.3*(e3)$) {$\mathbf{(-+-)}$};
\node[blue!80!green,fill=white,opacity=0.,text opacity=1] at ($0.3*(e1)+0.3*(e2)-0.6*(e3)$) {$\mathbf{(++-)}$};

\node[black] at ($0*(e1)+1.05*(e2)-1.05*(e3)$) {\tiny \textbf{(0,1,-1)}};
\node[black] at ($1.16*(e1)+0*(e2)-1.16*(e3)$) {\tiny \textbf{(1,0,-1)}};
\node[black] at ($1.05*(e1)-1.05*(e2)-0*(e3)$) {\tiny \textbf{(1,-1,0)}};
\node[black] at ($0*(e1)-1.05*(e2)+1.05*(e3)$) {\tiny \textbf{(0,-1,1)}};
\node[black] at ($-1.16*(e1)+0*(e2)+1.16*(e3)$) {\tiny \textbf{(-1,0,1)}};
\node[black] at ($-1.05*(e1)+1.05*(e2)-0*(e3)$) {\tiny \textbf{(-1,1,0)}};

\coordinate (C) at ($-2.2*(e1)$);

\draw[fill=black] ($(C)-(0.6,0)$) circle (5pt);
\draw[fill=black] ($(C)+(0.6,0)$) circle (5pt);
\draw[ultra thick] ($(C)+(0.6,0.08)$) -- ($(C)-(0.6,-0.08)$);
\draw[ultra thick] ($(C)+(0.6,0)$) -- ($(C)-(0.6,0)$);
\draw[ultra thick] ($(C)+(0.6,-0.08)$) -- ($(C)-(0.6,0.08)$);

\draw[ thick,postaction={decorate, decoration={markings,
   mark=at position 0.8 with {\arrow{stealth}}}}] ($(C)-(0.7,0)$)--($(C)+(-0.3,0.5)$);

\draw[thick, style={decorate, decoration={snake,amplitude=0.6,segment length=2.2}}] 
($(C)-(0.5,0)$)--($(C)+(-0.1,0.5)$);

\draw[ thick,postaction={decorate, decoration={markings,
   mark=at position 0.5 with {\arrow{stealth}}}}] ($(C)+(0.3,0.5)$)--($(C)+(0.7,0.)$);

\coordinate (C) at ($2*(e2)$);
\draw[white,fill=white] (C) circle (6pt);

\draw[fill=black] ($(C)-(0.6,0)$) circle (5pt);
\draw[fill=black] ($(C)+(0.6,0)$) circle (5pt);
\draw[ultra thick] ($(C)+(0.6,0.08)$) -- ($(C)-(0.6,-0.08)$);
\draw[ultra thick] ($(C)+(0.6,0)$) -- ($(C)-(0.6,0)$);
\draw[ultra thick] ($(C)+(0.6,-0.08)$) -- ($(C)-(0.6,0.08)$);

\draw[ thick,postaction={decorate, decoration={markings,
   mark=at position 0.8 with {\arrow{stealth}}}}] ($(C)+(0.5,0.)$)--($(C)+(0.1,0.5)$);

\draw[thick, style={decorate, decoration={snake,amplitude=0.6,segment length=2.2}}] 
($(C)-(0.7,0)$)--($(C)+(-0.3,0.5)$);

\draw[ thick,postaction={decorate, decoration={markings,
   mark=at position 0.5 with {\arrow{stealth}}}}] ($(C)+(0.3,0.5)$)--($(C)+(0.7,0.)$);
   
\coordinate (C) at ($-2.2*(e3)$);

\draw[fill=black] ($(C)-(0.6,0)$) circle (5pt);
\draw[fill=black] ($(C)+(0.6,0)$) circle (5pt);
\draw[ultra thick] ($(C)+(0.6,0.08)$) -- ($(C)-(0.6,-0.08)$);
\draw[ultra thick] ($(C)+(0.6,0)$) -- ($(C)-(0.6,0)$);
\draw[ultra thick] ($(C)+(0.6,-0.08)$) -- ($(C)-(0.6,0.08)$);

\draw[ thick,postaction={decorate, decoration={markings,
   mark=at position 0.5 with {\arrow{stealth}}}}] ($(C)+(-0.3,0.5)$)--($(C)-(0.7,0)$);

\draw[thick, style={decorate, decoration={snake,amplitude=0.6,segment length=2.2}}] 
($(C)-(0.5,0)$)--($(C)+(-0.1,0.5)$);

\draw[ thick,postaction={decorate, decoration={markings,
   mark=at position 0.8 with {\arrow{stealth}}}}] ($(C)+(0.7,0.)$)--($(C)+(0.3,0.5)$);

\coordinate (C) at ($2.2*(e1)$);
\draw[white,fill=white] (C) circle (8pt);

\draw[fill=black] ($(C)-(0.6,0)$) circle (5pt);
\draw[fill=black] ($(C)+(0.6,0)$) circle (5pt);
\draw[ultra thick] ($(C)+(0.6,0.08)$) -- ($(C)-(0.6,-0.08)$);
\draw[ultra thick] ($(C)+(0.6,0)$) -- ($(C)-(0.6,0)$);
\draw[ultra thick] ($(C)+(0.6,-0.08)$) -- ($(C)-(0.6,0.08)$);

\draw[ thick,postaction={decorate, decoration={markings,
   mark=at position 0.8 with {\arrow{stealth}}}}] ($(C)+(0.7,0)$)--($(C)-(-0.3,-0.5)$);

\draw[thick, style={decorate, decoration={snake,amplitude=0.6,segment length=2.2}}] 
($(C)+(0.5,0)$)--($(C)+(0.1,0.5)$);

\draw[ thick,postaction={decorate, decoration={markings,
   mark=at position 0.5 with {\arrow{stealth}}}}] ($(C)-(0.3,-0.5)$)--($(C)-(0.7,0.)$);

\coordinate (C) at ($-2*(e2)$);

\draw[fill=black] ($(C)-(0.6,0)$) circle (5pt);
\draw[fill=black] ($(C)+(0.6,0)$) circle (5pt);
\draw[ultra thick] ($(C)+(0.6,0.08)$) -- ($(C)-(0.6,-0.08)$);
\draw[ultra thick] ($(C)+(0.6,0)$) -- ($(C)-(0.6,0)$);
\draw[ultra thick] ($(C)+(0.6,-0.08)$) -- ($(C)-(0.6,0.08)$);

\draw[ thick,postaction={decorate, decoration={markings,
   mark=at position 0.8 with {\arrow{stealth}}}}] ($(C)-(0.7,0.)$)--($(C)-(0.3,-0.5)$);

\draw[thick, style={decorate, decoration={snake,amplitude=0.6,segment length=2.2}}] 
($(C)+(0.7,0)$)--($(C)+(0.3,0.5)$);


\draw[ thick,postaction={decorate, decoration={markings,
   mark=at position 0.5 with {\arrow{stealth}}}}] ($(C)+(-0.1,0.5)$)--($(C)-(0.5,0)$);
   
\coordinate (C) at ($2.2*(e3)$);
\draw[white,fill=white] (C) circle (8pt);

\draw[fill=black] ($(C)-(0.6,0)$) circle (5pt);
\draw[fill=black] ($(C)+(0.6,0)$) circle (5pt);
\draw[ultra thick] ($(C)+(0.6,0.08)$) -- ($(C)-(0.6,-0.08)$);
\draw[ultra thick] ($(C)+(0.6,0)$) -- ($(C)-(0.6,0)$);
\draw[ultra thick] ($(C)+(0.6,-0.08)$) -- ($(C)-(0.6,0.08)$);

\draw[ thick,postaction={decorate, decoration={markings,
   mark=at position 0.5 with {\arrow{stealth}}}}] ($(C)+(0.3,0.5)$)--($(C)+(0.7,0)$);

\draw[thick, style={decorate, decoration={snake,amplitude=0.6,segment length=2.2}}] 
($(C)+(0.5,0)$)--($(C)+(0.1,0.5)$);

\draw[ thick,postaction={decorate, decoration={markings,
   mark=at position 0.8 with {\arrow{stealth}}}}] ($(C)-(0.7,0.)$)--($(C)-(0.3,-0.5)$);

\end{tikzpicture}
\caption{Support domain for the twist-3 quark-gluon-quark PDFs (\ref{domain-x}) in the barycentric coordinates. Each sector is labeled by the signs of $(x_1,x_2,x_3)$. Each little diagram represents the corresponding partonic process. In these diagrams, the fields attached to each left circle are emitted, and the fields attached to each right circle are absorbed.}
\label{fig_partonic_regions}    
\end{center}
\end{figure}
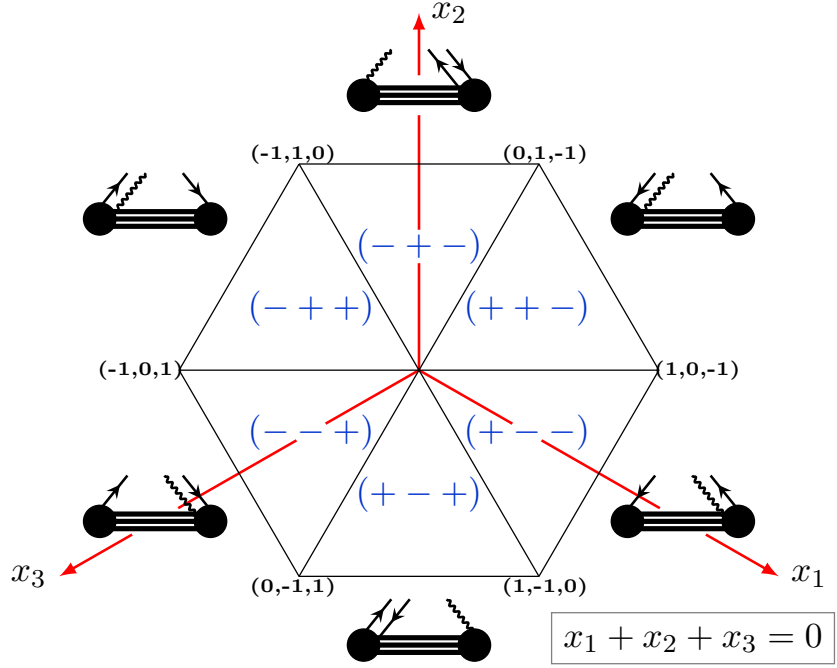

We start with a review of the properties of twist-3 PDFs that are important for the present paper. The twist-3 PDFs are defined as forward matrix elements of twist-3 operators. The latter can be presented as a T-ordered product of three fields positioned along the light-cone direction specified by a vector $n^\mu$ ($n^2 = 0$). The conventional momentum-fraction representation for a generic twist-3 PDF is obtained by the Fourier transform
\begin{align}
& \langle p,S|\mathcal{O}(z_1,z_2,z_3)|p,S\rangle\sim   \int [dx] e^{-ip^+\sum x_i z_i}f(x_1,x_2,x_3),
\end{align}
where $\mathcal{O}$ is the twist-3 operator with fields positioned at $(z_1 n, z_2 n, z_3 n)$, $f$ is the corresponding twist-3 PDF, and $x_i$ are (dimensionless) momentum fractions. The integration measure $[dx]$ is given below in  eqn.\eqref{momentum_integration_measure}. Causality and the momentum conservation endorse restrictions on the support of PDFs \cite{Jaffe:1983hp}:
\[\label{domain-x}
-1\le x_i \le 1,\qquad x_1+x_2+x_3=0.
\]
These restrictions are reflected in the integration measure as
\[
\label{momentum_integration_measure}
\int [dx] = \int_{-1}^1 dx_1 \int_{-1}^1 dx_2 \int_{-1}^1 dx_3 \delta(x_1+x_2+x_3).
\]
Therefore, a twist-3 PDF is a function of two independent variables. However, it is convenient to keep the three-variable notation, and visualise the PDFs in the barycentric coordinates, where their support takes the shape of a regular hexagon, see fig. \ref{fig_partonic_regions}. Twist-3 PDFs obey various symmetries with respect to reflection and rotation of the hexagon, which are summarized in secs.\ref{sec_qgq_pdfs} and \ref{sec_ggg_pdfs}.

Each sector of the hexagon domain corresponds to a different combination of signs of arguments $x_i$. Each combination defines a distinct elementary partonic processes, which in turn, provides the partonic interpretation of PDF. For example,  the subprocesses for a quark-gluon-quark PDF (i.e. $\mathcal{O}\sim \bar q F q$ as in (\ref{def:T} - \ref{def:HE})) are characterized by:
\begin{align*}
&\begin{cases}
x_1>0 & \text{emission of anti-quark,}\\ 
x_1<0 & \text{absorption of quark,}
\end{cases} \\
&\begin{cases}
x_2>0 & \text{emission of gluon,}\\ 
x_2<0 & \text{absorption of gluon,}
\end{cases} \\ 
&\begin{cases}
x_3>0 & \text{emission of quark,}\\ 
x_3<0 & \text{absorption of anti-quark.}
\end{cases}
\end{align*}
The momentum conservation condition $x_1+x_2+x_3=0$ prevents one from having all three signs positive (which would correspond to simultaneous emission of quark, anti-quark and gluon) or negative (simultaneous absorption of quark, anti-quark and gluon). A graphical representation of the support and region subdivision is given in Fig. \ref{fig_partonic_regions}.

In addition to three momentum-fractions, each PDF depends on the scale argument (conventionally denoted by $\mu$), which we omit in the majority of formulas for simplicity. This argument represents the physical scale of measurement of PDFs. The dependence on $\mu$ is dictated by evolution equations that are presented in sec.\ref{sec:evol}.

There are four independent quark-gluon-quark PDFs, and two independent gluon-gluon-gluon PDFs (for simplicity we call them quark and gluon PDFs). Their definitions are presented below. Note that there is no standardized notation and parametrization for twist-3 PDFs. The main reason is that twist-3 PDFs are involved functions, and there is no notation that would be convenient in all applications. Also, despite the discovery of twist-3 physics long ago, there have been few practical attempts to operate with it. Therefore, different groups often invent their own notation.

In particular, the evolution equations are simplified in the so-called definite-C-parity basis, which we use for the numerical computation. However, practically, one faces combinations of functions that possess different symmetries and properties than the definite-C-parity basis. Thus, for applications, it is advantageous to pass to the physics-motivated basis. The latter has no standard parametrization. Here, we adopt the parametrization from refs.\cite{Scimemi:2019gge, Braun:2021aon, Braun:2021gvv, Rein:2022odl}. The parametrizations of quark-gluon-quark matrix elements are rather similar between different studies, see for instance \cite{Kang:2008ey, Boer:1997bw, Kanazawa:2000hz, Boer:2003cm, Eguchi:2006mc, Braun:2009mi, Scimemi:2018mmi}. The parametrization for the three-gluon matrix elements could differ significantly between different works, compare \cite{Braun:2000yi, Kang:2008ey, Beppu:2010qn, Scimemi:2019gge}. In the following sections, we provide a translation dictionary to some other parametrizations. 

\subsection{Quark-gluon-quark PDFs}
\label{sec_qgq_pdfs}
We define the quark PDFs of twist-3 as:
\begin{eqnarray}\label{def:T}
&& \langle p,S|g\bar q(z_1n)F^{\mu+}(z_2n) \gamma^+q(z_3n)|p,S\rangle  
\\ \notag &&
\qquad = 2 \epsilon^{\mu\nu}_T s_\nu (p^+)^2 M \int[dx] e^{-ip^+(x_1z_1+x_2z_2+x_3z_3)}T(x_{123}),
\\\label{def:deltaT}
&&\langle p,S|ig\bar q(z_1n)F^{\mu+}(z_2n) \gamma^+\gamma^5q(z_3n)|p,S\rangle  
\\\notag &&
\qquad
= -2 s_T^\mu (p^+)^2 M \int[dx] e^{-ip^+(x_1z_1+x_2z_2+x_3z_3)}\Delta T(x_{123}),
\\\label{def:HE}
&&\langle p,S|g\bar q(z_1n)F^{\mu+}(z_2n) i\sigma^{\nu+}\gamma^5q(z_3n)|p,S\rangle  
\\\notag &&
\qquad
= 2 (p^+)^2 M \int[dx] e^{-ip^+(x_1z_1+x_2z_2+x_3z_3)}
\qa \epsilon^{\mu\nu}_T E(x_{123})+is_L g^{\mu\nu}_T H(x_{123})\qc, \notag
\end{eqnarray}
where $(x_{123})$ is a short-hand notation for $(x_1,x_2,x_3)$, $g$ is the QCD coupling constant, $q$($\bar q$) is the (anti-)quark field and $F^{\mu\nu}$ is the gluon field-strength tensor. The vector $s^\mu$ is the spin-vector, and $s_L$ is its longitudinal component. The vector $n$ is a light-cone vector
\footnote{
We use the standard convention for the notation of component of a vector in the light-cone decomposition. For a generic vector $a$ one has
\[
\notag
a^\mu = a^+ \bar{n}^\mu + a^- n^\mu + a_T^\mu
\]
where $(n\bar{n})\equiv n_\mu \bar{n}^\mu = 1$ and $n^2 = \bar{n}^2 = (na_T) = (\bar{n}a_T) = 0$ from which $a^+\equiv (na)$ and $a^- = (\bar{n}a)$.}.
All open indices are transverse, and
\begin{eqnarray}
\label{Metric_and_epsilon_def}
g_T^{\mu\nu}=g^{\mu\nu}-n^\mu \bar n^\nu-\bar n^\mu n^\nu,\qquad \epsilon^{\mu\nu}_T=\epsilon^{\mu\nu-+}, \quad \e^{0123} = 1.
\end{eqnarray}
In formulas (\ref{def:T}-\ref{def:HE}) we have omitted the straight gauge-links that connect the fields for simplicity. These links should be inserted into each operator and contracted at any position. For instance
\[
\begin{split}
&\bar{q}_i(z_1n)[z_1n,z_2n]F^{\mu+}(z_2n)[z_2n,z_3n]q(z_3n) 
\\
& = \bar{q}_i(z_1n)[z_1n,z_0n]_{ii_0} F_{ln}^{\mu+}(z_2n)[z_2n,z_0n]_{l_0l}[z_2n,z_0n]_{nn_0} [z_0n,z_3n]_{j_0j}q(z_3n) \delta_{i_0l_0}\delta_{n_0j_0}
\end{split}
\]
where $F_{ln}^{\mu+} \equiv T^a_{ln} F^{\mu+}_a$, and similarly for the other color structures. Gauge links ensure the gauge-invariance of the operators. The present parametrization is adopted from \cite{Braun:2009mi}\footnote{We note that in the last line of  eqn.(5) of \cite{Braun:2009mi} the $\tilde{s}^\mu$ should be a $-s^\mu$ . Also, ref.\cite{Braun:2009mi} uses a different notation which includes a mass factor in the distributions:
\begin{align*} 
T_{\bar{q}Fq \text{\cite{Braun:2009mi}}}(x_{123}) &= MT(x_{123}),\\
\Delta T_{\bar{q}Fq \text{\cite{Braun:2009mi}}}(x_{123}) &= M\Delta T(x_{123}).
\end{align*}
} (for chiral-even distributions) and \cite{Ji:2014eta,Braun:2021gvv} (for chiral-odd).

Thus, there are four quark PDFs of twist-3, denoted as $T$, $\Delta T$, $H$, and $E$. The PDFs $T$ and $\Delta T$ are chiral-even, while $H$ and $E$ are chiral-odd. The chiral pairs of distributions ($T$, $\Delta T$) and ($H$, $E$) appear in different processes. The pairs cannot mix with the other, because the chirality is preserved in massless QCD. 

The functions $T$ and $\Delta T$ do not have definite charge-conjugation parity. Meanwhile, QCD evolution preserves that symmetry, and thus the corresponding evolution equations are simpler for parity-even and -odd components. Following ref.\cite{Braun:2009mi}, we introduce definite-C-parity functions $\mathfrak{S}$ as
\[
\begin{split}
\label{frakS->T}
T(x_{123}) &= \frac{1}{4} \qa \mathfrak{S}^+(x_{123})+\mathfrak{S}^+(x_{321})+\mathfrak{S}^-(x_{123})-\mathfrak{S}^-(x_{321})\qc,\\
\Delta T(x_{123}) &= -\frac{1}{4} \qa \mathfrak{S}^+(x_{123})-\mathfrak{S}^+(x_{321})+\mathfrak{S}^-(x_{123})+\mathfrak{S}^-(x_{321})\qc.
\end{split}
\]
The order of indices designates the permutation of arguments, for instance, $(x_{321})$ corresponds to $(x_3,x_2,x_1)$. The chiral-odd distributions $H$ and $E$ have definite charge-conjugation parity, and thus there is no need for a change the basis. 

Another conventional composition is 
\begin{eqnarray}
\label{S->T}
T(x_{123})=- S^+(x_{123})- S^-(x_{123}),\qquad
\Delta T(x_{123})=S^+(x_{123})- S^-(x_{123}).
\end{eqnarray}
The functions $S^\pm$ naturally appear in the description of structure functions \cite{Kodaira:1994ge, Braun:2001qx}, and two-point correlators \cite{Braun:2021aon}. The inverse transformations read
\begin{eqnarray}\label{frakS<-T<-S}
\mathfrak{S}^\pm(x_{123}) &=& T(x_{123})-\Delta T(x_{123}) \pm T(x_{321}) \pm \Delta T(x_{321}) \\
\notag &=&-2(S^+(x_{123})\pm S^-(x_{321})).
\\\label{S+<-frakS<-T}
S^+(x_{123}) &=& \frac{-T(x_{123})+\Delta T(x_{123})}{2}=-\frac{\mathfrak{S}^+(x_{123})+\mathfrak{S}^-(x_{123})}{4},
\\\label{S-<-frakS<-T}
S^-(x_{123}) &=&
\frac{-T(x_{123})-\Delta T(x_{123})}{2}=
-\frac{\mathfrak{S}^+(x_{321})-\mathfrak{S}^-(x_{321})}{4}.
\end{eqnarray}
With the help of these relations, one transforms twist-3 PDFs from one basis to another.

The twist-3 PDFs obey symmetries with respect to permutations of arguments. These symmetries follow from the discrete symmetries of QCD (see e.g. \cite{Braun:2009mi, Scimemi:2019gge}). In particular, the distributions $T$,  $\Delta T$, $H$ and $E$ obey the relations
\begin{align}\label{symm:T-quark}
&T(x_{123}) = T(-x_{321}), \quad \Delta T(x_{123}) = -\Delta T(-x_{321}),\\
\label{symm:HE}
&E(x_{123}) = E(-x_{321}), \quad H(x_{123}) = -H(-x_{321}),
\end{align}
here $(-x_{321})$ stays for the argument $(-x_3,-x_2,-x_1)$. The definite-C-parity distributions $\mathfrak{S}^\pm$ obey
\begin{eqnarray}
\label{symm:frakS-quark}
\mathfrak{S}^\pm(x_1,x_2,x_3) = \pm \mathfrak{S}^\pm(-x_1,-x_2,-x_3).
\end{eqnarray}
While PDFs $S^\pm$ obey
\begin{eqnarray}\label{symm:S-quark}
S^\pm(x_{123})=S^\mp(-x_{321}).
\end{eqnarray}
These symmetry properties are important to fulfill during the specification of the boundary conditions for the evolution equations.

\subsection{Three-gluon PDFs}\label{sec_ggg_pdfs}
We start by defining the three-gluon PDFs using an over-complete basis of functions:
\begin{eqnarray}
&& \langle p,S|igf^{ABC}F^{\mu+}_A(z_1n)F^{\nu+}_B(z_2n) F^{\rho+}_C(z_3n)|p,S\rangle  
\\\notag &&
\qquad\qquad
= 
(p^+)^3 M \int[dx] e^{-ip^+(x_1z_1+x_2z_2+x_3z_3)}\sum_{i=2,4,6}t_i^{\mu\nu\rho}F_i^+(x_{123}),
\\
&& \langle p,S|gd^{ABC}F^{\mu+}_A(z_1n)F^{\nu+}_B(z_2n) F^{\rho+}_C(z_3n)|p,S\rangle  
\\\notag &&
\qquad\qquad
= 
(p^+)^3 M \int[dx] e^{-ip^+(x_1z_1+x_2z_2+x_3z_3)}\sum_{i=2,4,6}t_i^{\mu\nu\rho}F_i^-(x_{123}),
\end{eqnarray}
where $f^{ABC}$ and $d^{ABC}$ are anti-symmetric and symmetric structure constants of SU($N_c$) algebra. The tensors $t$ are defined
\begin{eqnarray}\notag
t_2^{\mu\nu\rho}&=&
s^\alpha_T \epsilon^{\mu\alpha}_T g_T^{\nu\rho}
+s^\alpha_T \epsilon^{\nu\alpha}_T g_T^{\rho\mu}
+s^\alpha_T \epsilon^{\rho\alpha}_T g_T^{\mu\nu},
\\\label{def:tensor-t}
t_4^{\mu\nu\rho}&=&
-s^\alpha_T \epsilon^{\mu\alpha}_T g_T^{\nu\rho}
+2s^\alpha_T \epsilon^{\nu\alpha}_T g_T^{\rho\mu}
-s^\alpha_T \epsilon^{\rho\alpha}_T g_T^{\mu\nu},
\\\notag
t_6^{\mu\nu\rho}&=&
s^\alpha_T \epsilon^{\mu\alpha}_T g_T^{\nu\rho}
-s^\alpha_T \epsilon^{\rho\alpha}_T g_T^{\mu\nu}.
\end{eqnarray}
Note that these tensors are defined in 4-dimensions. In the following, we consider only the 4-dimensional case. In D-dimensions, there are 6 possible tensors and hence 6 PDFs (see the complete decomposition in appendix A of ref.\cite{Scimemi:2019gge}). Additionally, symmetry relations for functions are more complicated in D-dimensions.

The functions $F^\pm_{2,4,6}$ obey a large number of symmetries and relations that derive from the symmetries of the operators under the permutation of fields. The application of these symmetries reduces the number of independent functions and results in alternative conventions of PDF basis. The popular choice for the basis is four functions ($T_{3F}^\pm$, $\Delta T_{3F}^\pm$) introduced in refs.\cite{Braun:2000yi, Braun:2009mi}. They are defined as\footnote{We notice that in \cite{Braun:2000yi, Braun:2009mi} the convention for the anti-symmetric tensor is $\e_{0123} = 1$, which is opposite to the convention used here, see  eqn.\eqref{Metric_and_epsilon_def}.}
\begin{align}
\label{braun_to_us_T3F}
T_{3F}^\pm(x_{123}) & = 4F_2^\pm(x_{123}) + 2F_4^\pm(x_{123}), \\
\label{braun_to_us_DT3F}
\Delta T_{3F}^\pm(x_{123}) & = -2F_6^\pm(x_{123}).
\end{align}
These four functions can be reduced to two independent definite-C-parity PDFs $\mathfrak{F}^\pm$,
\begin{eqnarray}\label{T=F}
T^{\pm}_{3F}(x_{123}) &=& \frac{1}{2}\qa  \mathfrak{F}^{\pm}(x_{123})\mp \mathfrak{F}^{\pm}(x_{321})\qc,\\
\label{deltaT=F}
\Delta T^{\pm}_{3F}(x_{123}) &=& -\frac{1}{2}\qa  \mathfrak{F}^{\pm}(x_{123})\pm \mathfrak{F}^{\pm}(x_{321})\qc.    
\end{eqnarray}
Note that PDFs $T_{3F}^\pm$ and $\Delta T_{3F}^\pm$ are not independent, but satisfy
\begin{eqnarray}
\Delta T_{3F}^\pm(x_{123}) & =& \pm \qa T^\pm_{3F}(x_{132}) - T_{3F}^\pm(x_{213})\qc.
\end{eqnarray}
The inverse relations read
\begin{align}\label{F<-T(gluon)}
\mathfrak{F}^\pm(x_{123}) &= T^\pm_{3F}(x_{123}) - \Delta T_{3F}^\pm (x_{123}) =  T^\pm_{3F}(x_{123}) \mp T^\pm_{3F}(x_{132}) \pm T_{3F}^\pm(x_{213})
\end{align}
The definite-C-parity PDFs $\mathfrak{F}^\pm$ do not mix in the evolution. Thus, we utilize them as the basis for the evolution. However, in practice one often refers to $F$-basis or $T$-basis.

Three-gluon PDFs have a very rich set of symmetries:
\begin{eqnarray}\label{symm:T-gluon}
T_{3F}^\pm(x_{123}) &=&  T_{3F}^\pm(-x_{321}), \qquad \qquad\qquad T_{3F}^\pm(x_{123}) = \mp T_{3F}^\pm(x_{321}),
\\
\Delta T_{3F}^\pm(x_{123}) &=&  -\Delta T_{3F}^\pm(-x_{321}),
 \qquad \qquad \Delta T_{3F}^\pm(x_{123}) = \pm \Delta T_{3F}^\pm(x_{321}),
\\\label{symm:F-gluon}
\mathfrak{F}^\pm(x_1,x_2,x_3) &=& \mp \mathfrak{F}^\pm(-x_1,-x_2,-x_3), \quad \mathfrak{F}^\pm(x_1,x_2,x_3) = \mp \mathfrak{F}^\pm(x_1,x_3,x_2).
\end{eqnarray}
Similarly, to the quark case these symmetries must be satisfied by boundary conditions.

Different parametrizations are natural (or convenient) when studying different aspects of the parton physics. Another convenient parametrization has been suggested in \cite{Scimemi:2019gge}. It includes two functions, which have very rich symmetries \footnote{
In refs.\cite{Rein:2022odl,Scimemi:2019gge} the gluon PDFs $G^\pm$ and $Y^\pm$ are defined. They are related to distributions $T_{3F}^\pm$ and $\Delta T_{3F}^\pm$ as
\begin{align}\nn
T^\pm_{3F}(x_{123}) &
= G^\pm(x_{123})+Y^\pm(x_{123}),
\qquad\qquad
\Delta T^\pm_{3F}(x_{123}) 
= \mp \ta Y^\pm(x_{132})-Y^\pm(x_{213})\tc .
\end{align}
The symmetries of $G^\pm$, and $Y^\pm$ are
\begin{align*}
G_\pm(x_{123}) &= G_\pm(-x_{321}) = \mp G_\pm(x_{213}) = \mp G_\pm(x_{132})\\
Y_\pm(x_{123}) &= Y_\pm(-x_{321}) = \mp Y_\pm(x_{321}), \\
0&=Y_\pm(x_{123}) + Y_\pm(x_{231}) + Y_\pm(x_{312}).
\end{align*}
These functions form convenient combinations in the perturbative computations related to transverse momentum dependent processes.
}.
Meanwhile, it is difficult to parameterize functions with cumbersome symmetries. Therefore, in the present implementation we restrict ourselves to the functions $T_{3F}^\pm$ or $\mathfrak{F}^\pm$ as the input.

\subsection{Quark threshold and flavor decomposition}

The quark operator can be constructed with quark fields of different flavor (both fields of the same flavor). Depending on the energy, one can find $u$, $d$, $s$, $c$ and $b$ quark contributions (there is also $t$ quark but its contribution is neglected, since it is very heavy). The number of active quark flavors is dependent on the energy scale. In the present implementation we use the simplest Zero Mass Variable Flavour Number Scheme (ZMVFNS) \cite{Collins:1986mp}. This scheme consists of neglecting parton distribution below a certain threshold and treating all active quarks as massless. Thus, we select two threshold scales\footnote{The value of threshold scales can be specified within the code, but the hierarchy $\mu_c<\mu_b$ should be preserved.}
\begin{eqnarray}
\mu_c\sim m_c\sim 1.3\text{GeV},\qquad \mu_b\sim m_b\sim 4.2\text{GeV}.
\end{eqnarray}
The $u$, $d$ and $s$ quarks are always included. Evolution below for scale $\mu<\mu_c$ is computed accounting only for $u$, $d$ and $s$ contributions with $c$ and $b$ PDFs being set to zero. For $\mu_c<\mu<\mu_b$ also the contribution of $c$-quark is accounted, and for $\mu>\mu_b$ also the contribution of $b$-quark is accounted. In other words we use the following setting
\begin{eqnarray}\nn
\mu<\mu_c &:& n_f=3~,\\
\mu_c<\mu<\mu_b &:& n_f=4~,\\\nn
\mu_b<\mu &:& n_f=5~,
\end{eqnarray}
where $n_f$ designates the number of active flavors.

At the threshold scales $\mu=\mu_c$ and $\mu=\mu_b$ the values of PDFs are equalized. Practically, it is implemented as three independent evolution procedures which use values at thresholds as boundary conditions. Importantly, such a procedure is consistent only for the evolution solution from low to high scales.

Along the evolution process, PDFs of different flavors mix with (the total sum of) each other and with the gluon PDF. One can select particular combinations of quark PDFs (non-singlets) that evolve independently, whereas the singlet contribution mixes with gluon. Note that in the case of twist-2 PDFs, one selects autonomous contribution simply considering the difference between the quark and anti-quark distributions (with positive momentum fractions). In the twist-3 case it is not possible, since there is no natural notion of anti-quark distribution that is preserved by evolution. Instead, we define
\begin{eqnarray}
\text{singlet}=\mathfrak{S}^\pm_{\text{\tiny S}} = \sum_{f}\mathfrak{S}^\pm_{f},
\end{eqnarray}
where index $f$ indicates the flavor, and sum includes $n_f$ active quarks. The non-singlet combinations are
\begin{eqnarray}
\mathfrak{S}^\pm_{\text{\tiny NS}_1} &=& \mathfrak{S}^\pm_{u}-\mathfrak{S}^\pm_{d}, 
\\
\mathfrak{S}^\pm_{\text{\tiny NS}_2}  &=& \mathfrak{S}^\pm_{u}+\mathfrak{S}^\pm_{d}-2\mathfrak{S}^\pm_{s},
\\
\mathfrak{S}^\pm_{\text{\tiny NS}_3}  &=& \mathfrak{S}^\pm_{u}+\mathfrak{S}^\pm_{d}+\mathfrak{S}^\pm_{s}-3\mathfrak{S}^\pm_{c},
\\
\mathfrak{S}^\pm_{\text{\tiny NS}_4}  &=& \mathfrak{S}^\pm_{u}+\mathfrak{S}^\pm_{d}+\mathfrak{S}^\pm_{s}+\mathfrak{S}^\pm_{c}-4\mathfrak{S}^\pm_{b}.
\end{eqnarray}
At quark-mass thresholds the definition of singlet changes, since $n_f$ increases.
\begin{eqnarray}
\mathfrak{S}^\pm_{\text{\tiny S}}(\mu_{\text{thr.}})|_{n_f}=\mathfrak{S}^\pm_{\text{\tiny S}}(\mu_{\text{thr.}})|_{n_f+1},
\end{eqnarray}
where $n_f$ is the number of active quarks below $\mu_{\text{thr.}}$. Also, non-singlet combinations $\mathfrak{S}^\pm_{\text{\tiny NS}_3}$ and $\mathfrak{S}^\pm_{\text{\tiny NS}_4}$ do not exist below $\mu_c$ and $\mu_b$ correspondingly, whereas at thresholds they read
\begin{eqnarray}
\mathfrak{S}^\pm_{\text{\tiny NS}_3}(\mu_c)=\mathfrak{S}^\pm_{\text{\tiny S}}(\mu_c),\qquad
\mathfrak{S}^\pm_{\text{\tiny NS}_4}(\mu_b)=\mathfrak{S}^\pm_{\text{\tiny S}}(\mu_b).
\end{eqnarray}
Within the code the evolution equations are solved for singlet and non-singlet combinations, while the boundary input and output are provided for individual flavors.

There is no necessity to decompose chiral-odd distributions $H$ and $E$, because they do not mix with each other and/or gluon.

\subsection{Evolution equation for twist-three PDFs}
\label{sec:evol}

The evolution equations for chiral-even twist-3 PDFs have the following structure:
\begin{align}\label{ev-1}
\mu^2 \frac{\partial}{\partial \mu^2}\mathfrak{S}_{\text{\tiny NS}_i}^\pm &= -a_s(\mu) \mathbb{H}_{\text{\tiny NS}}\mathfrak{S}_{\text{\tiny NS}_i}^\pm 
~,
\\
\mu^2 \frac{\partial}{\partial \mu^2} \begin{pmatrix}
    \mathfrak{S}_S^\pm \\
    \mathfrak{F}^\pm
\end{pmatrix} &= -a_s(\mu) \begin{pmatrix}
    \mathbb{H}^\pm_{qq} & \mathbb{H}^\pm_{qg} \\
    \mathbb{H}^\pm_{gq} & \mathbb{H}^\pm_{gg} \\
\end{pmatrix}\begin{pmatrix}
    \mathfrak{S}_S^\pm \\
    \mathfrak{F}^\pm
\end{pmatrix},
\end{align}
where $a_s(\mu)$ is the QCD coupling constant ($a_s=g^2/(4\pi)^2$), and $\mathbb{H}$ are integral evolution kernels. The arguments of PDFs $(x_1,x_2,x_3;\mu)$ are omitted on both sides. The notation $\mathbb{H}f$ implies that the kernel is convoluted with the function. The chiral-odd distributions evolve according to the equations
\begin{eqnarray}\label{ev-3}
\mu^2 \frac{\partial}{\partial \mu^2} E &=& -a_s(\mu) \mathbb{H}_{\text{\tiny CO}}E, 
\\\nn 
\mu^2 \frac{\partial}{\partial \mu^2} H&=& -a_s(\mu) \mathbb{H}_{\text{\tiny CO}}H.
\end{eqnarray}

The non-singlet chiral-even, chiral-odd and diagonal parts of the singlet kernels are conveniently written in terms of the two-particle $SL(2)$-invariant operators \cite{Braun:2009vc, Braun:2009mi}:
\begin{align}
\mathbb{H}_{\text{\tiny NS}} &= N_c \ta \widehat{\mathcal{H}}_{12}+\widehat{\mathcal{H}}_{23}-2\mathcal{H}^+_{12}\tc  - \frac{1}{N_c} \ta\widehat{\mathcal{H}}_{13}-\mathcal{H}_{13}^+-\mathcal{H}_{23}^{e,(1)}P_{23}+2\mathcal{H}^-_{12}\tc - 3C_F \mathbf{1},\\
\mathbb{H}_{\text{\tiny CO}} &= N_c \ta \widehat{\mathcal{H}}_{12}+\widehat{\mathcal{H}}_{23} - 2\mathcal{H}^+_{12}-2\mathcal{H}^+_{23}\tc -\frac{1}{N_c} \ta \widehat{\mathcal{H}}_{13} + 2\mathcal{H}^-_{12} + 2\mathcal{H}^-_{23}\tc-3C_F \mathbf{1}\\ 
\mathbb{H}^+_{qq} &= \mathbb{H}_{\text{\tiny NS}} + 4n_f \mathcal{H}^d_{13}, \quad \mathbb{H}^-_{qq} = \mathbb{H}_{\text{\tiny NS}},\\
\mathbb{H}^\pm_{gg} &= N_c \Big[\widehat{\mathcal{H}}_{12}+\widehat{\mathcal{H}}_{23}+\widehat{\mathcal{H}}_{31} -4\ta \mathcal{H}^+_{12}+ \mathcal{H}^+_{13}\tc -2 \ta \widetilde{\mathcal{H}}^+_{12}+ \widetilde{\mathcal{H}}^+_{13}\tc  \\
& \notag \qquad\quad \pm 6\ta\mathcal{H}^-_{12}+ \mathcal{H}^-_{13}\tc\Big] - \beta_0 \mathbf{1},
\end{align}
where $\mathbf{1}$ is the unity convolution, $N_c$ is the number of colors ($=3$), $C_F=(N_c^2-1)/(2N_c)(=4/3)$ and 
\[
\beta_0 = \frac{11N_c-2n_f}{3}.
\]

Instead, the non-diagonal entries of the singlet kernel, albeit having their own expressions in terms of $SL(2)$-invariant kernels, are more conveniently presented in a specialized form
\begin{align}
\mathbb{H}^\pm_{qg} &= n_f \ta \mathcal{V}_{13}^+ \mp \mathcal{V}_{13}^-\tc, \\
\mathbb{H}^+_{gq} &= N_c (1-P_{23})\ta \mathcal{W}^++\mathcal{W}^--2\Delta \mathcal{W}\tc ,\\
\mathbb{H}^-_{gq} &= -\frac{N_c^2-4}{N_c} (1+P_{23})\ta \mathcal{W}^++\mathcal{W}^-\tc ,
\end{align}
where $P_{23}$ is the operation of permutation of arguments $x_2\leftrightarrow x_3$.
The explicit expressions for the kernels in momentum space are collected in \ref{sec_app_evolutionkernels}.

\begin{figure}[!ht]
\begin{center}
\begin{tikzpicture}
\tikzmath{\s = 1.5;}

\coordinate (e1) at (0.866025*\s, -0.5*\s);
\coordinate (e2) at (0.*\s, 1.*\s);
\coordinate (e3) at (-0.866025*\s, -0.5*\s);

\draw[white,fill=white] ($-2.5*(e2)$) circle (5pt);
\draw[white,fill=white] ($2.8*(e2)$) circle (5pt);

\draw[-latex,thick,grey] (0,0) -- ($3*(e1)$);
\draw[-latex,thick,grey] (0,0) -- ($2.6*(e2)$);
\draw[-latex,thick,grey] (0,0) -- ($3*(e3)$);

\node[right] at ($3*(e1)$) {$x_1$};
\node[right] at ($2.6*(e2)$) {$x_2$};
\node[left] at ($3*(e3)$) {$x_3$};

\draw[thick, dashed] ($-0.8*(e1)+0.8*(e2)+0*(e3)$) -- 
		($-0.8*(e1)+0*(e2)+0.8*(e3)$) --
		($0*(e1)-0.8*(e2)+0.8*(e3)$) --
		($0.8*(e1)-0.8*(e2)-0*(e3)$) --
		($0.8*(e1)+0*(e2)-0.8*(e3)$) --
		($0*(e1)+0.8*(e2)-0.8*(e3)$) --
		($-0.8*(e1)+0.8*(e2)+0*(e3)$);

\draw[very thick] ($-1.5*(e1)+1.5*(e2)+0*(e3)$) -- 
		($-1.5*(e1)+0*(e2)+1.5*(e3)$) --
		($0*(e1)-1.5*(e2)+1.5*(e3)$) --
		($1.5*(e1)-1.5*(e2)-0*(e3)$) --
		($1.5*(e1)+0*(e2)-1.5*(e3)$) --
		($0*(e1)+1.5*(e2)-1.5*(e3)$) --
		($-1.5*(e1)+1.5*(e2)+0*(e3)$);

\node[above] at ($-1.2*(e2) + 0.7*(e1)-0.7*(e2) + (0.2,0)$) {$\bm{(1)}$};
\draw[blue,ultra thick] ($-1.2*(e2)$) -- ($-1.2*(e2) + 0.7*(e1)-0.7*(e2)$);

\node[above] at ($-1.2*(e2) + 0.7*(e3)-0.7*(e2) + (-0.2,0)$) {$\bm{(2)}$};
\draw[red,ultra thick] ($-1.2*(e2)$) -- ($-1.2*(e2) + 0.7*(e3)-0.7*(e2)$);

\node[below] at ($-1.2*(e2) + 1.1*(e3)-1.1*(e1) + (+0.5,0)$) {$\bm{(3)}$};
\draw[myGold,ultra thick] ($-1.2*(e2)$) -- ($-1.2*(e2) + 1.1*(e3)-1.1*(e1)$);

\node[] at ($-1.2*(e1) - 0.7*(e1) + 0.7*(e2) + (+0.5,0)$) {$\bm{(4)}$};
\draw[cyan,ultra thick] ($-1.2*(e1)$) -- ($-1.2*(e1) - 0.7*(e1) + 0.7*(e2)$);

\node[] at ($-1.2*(e3) - 0.7*(e3) + 0.7*(e2) + (-0.5,0)$) {$\bm{(5)}$};
\draw[magenta,ultra thick] ($-1.2*(e3)$) -- ($-1.2*(e3) - 0.7*(e3) + 0.7*(e2)$);

\node[below] at ($-1.2*(e2) + 1.1*(e1)-1.1*(e3) + (-0.5,0)$) {$\bm{(6)}$};
\draw[myGreen,ultra thick] ($-1.2*(e2)$) -- ($-1.2*(e2) + 1.1*(e1)-1.1*(e3)$);

\draw[black,fill=black] ($-1.2*(e2)$) circle (3pt);



\end{tikzpicture}
\caption{Graphical representation of the integration path for the non-singlet kernel when the root point $x_1,x_2,x_3$ is in the $(+-+)$ region.}
\label{fig_integration_paths}    
\end{center}
\end{figure}
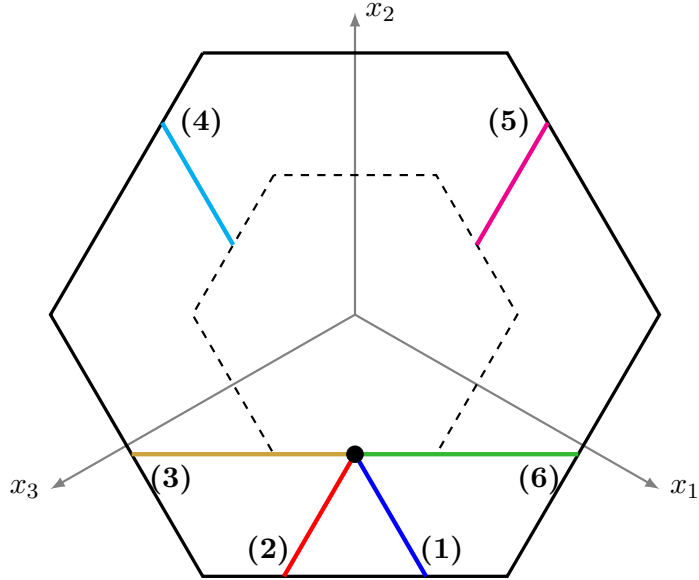

The elementary kernels have rather different analytical structures and expressions. Nonetheless, they preserve the following main property. For a given point $(x_1,x_2,x_3)$ the integral convolution involves only the ``outer'' region of support, which is a part of support that is complementary to the ``inner'' hexagon that passes through the point $(x_1,x_2,x_3)$ (see fig.\ref{fig_integration_paths}, where ``inner'' hexagon is drawn by dashed line). It is convenient to introduce the ``radius'' of the point $(x_1,x_2,x_3)$, as $r=\max(|x_1|,|x_2|,|x_3|)$. In this terminology, function values at radius $r$ are insensitive to the values at radii smaller than $r$, but sensitive all other values. It is a two-dimensional generalization of similar feature known for twist-2 evolution, where evolution for $f(x)$ involves only values in the range $[x,1]$. This property is a consequence of energy and momentum conservation and is preserved at all orders of perturbation theory.

At LO the evolution kernels have additional simplifications being reducible to line integrals. For example, the non-singlet evolution involves the following types of integrals (ignoring explicit expressions and focusing solely on the integration regions)
\begin{align}\label{multi-color}
& \mathbb{H}_{\text{\tiny NS}}\mathfrak{S}_{\text{\tiny NS}}^\pm \sim\\
\bm{(1)} \quad &\sim  {\color{blue}\int dv \qa \Theta(x_1,-v)f_1(v) + \Theta(x_2,v)f_2(v)\qc \mathfrak{S}_{\text{\tiny NS}}^{\pm}(x_1-v,x_2+v,x_3)} \notag \\
\bm{(2)} \quad & + {\color{red}\int dv \qa \Theta(x_3,-v)f_3(v) + \Theta(x_2,v)f_4(v)\qc \mathfrak{S}_{\text{\tiny NS}}^{\pm}(x_1,x_2+v,x_3-v)} \notag\\
\bm{(3)} \quad & + {\color{myGold} \int dv\Theta(x_3,v)f_5(v) \mathfrak{S}_{\text{\tiny NS}}^{\pm}(x_1-v,x_2,x_3+v)}  \notag\\
\bm{(4)} \quad & +{\color{cyan}\int dv \qa \Theta(x_1,-v)f_6(v) + \Theta(x_2,v)f_7(v)\qc \mathfrak{S}_{\text{\tiny NS}}^{\pm}(x_2+v,x_1-v,x_3)}\notag\\
\bm{(5)} \quad & +{\color{magenta}\int dv \Theta(x_3,-v)f_8(v)  \mathfrak{S}_{\text{\tiny NS}}^{\pm}(x_1,x_3-v,x_2+v)}\notag\\
\bm{(6)} \quad & +{\color{myGreen}\int dv  \Theta(x_1,-v)f_9(v)\mathfrak{S}_{\text{\tiny NS}}^{\pm}(x_1-v,x_2,x_3+v)} \notag, 
\end{align}
where $\Theta(a,b)=\theta(a)\theta(b)-\theta(-a)\theta(-b)$ is the combination of Heaviside theta functions. In fig. \ref{fig_integration_paths} we show the paths of integrals for individual contributions if the root point $(x_1,x_2,x_3)$ is in the $(+-+)$ region. The color-coding and numbering of paths coincide with the color-coding and numbering of the lines in eqn. (\ref{multi-color}). 
Under evolution, a change in the boundary conditions in any point of the region between the inner and outer hexagons in fig. \ref{fig_integration_paths}, has an impact on the value of the solution in any of the points of the inner hexagon.

\section{Discretization}
\label{sec_discretizaitons}

The code presented in this work solves the evolution equations (\ref{ev-1} - \ref{ev-3}) numerically starting from a specified boundary condition. The central part of the algorithm is the discretization procedure, which consists of the following. We specify a grid on the hexagon by mapping a point $\bm{x}=(x_1,x_2,x_3)$ that belongs to the hexagon to a pair of integers $(i,j)$. For a given grid one can approximate twist-3 PDF by
\begin{eqnarray}
\label{eq_global_mode_decomp}
\mathfrak{S}_{\text{\tiny NS}}(\bm{x})=\sum_{i,j} \mathcal{I}_{ij}(\bm{x})\mathfrak{S}_{\text{\tiny NS}}^{ij},\qquad \mathfrak{S}_{\text{\tiny NS}}^{ij}=\mathfrak{S}_{\text{\tiny NS}}(\bm{x}(i,j)),
\end{eqnarray}
where $\mathcal{I}$ is the interpolation function. 
In 2D there are multiple possible choices of how to interpolate the function, but any choice compatible with equation \eqref{eq_global_mode_decomp} must satisfy $\mathcal{I}_{ij}(\bm{x}(i',j'))=\delta_{ii'}\delta_{jj'}$. 

In the discrete approximation the evolution equation (\ref{ev-1}) can be written as
\begin{eqnarray}
\mu^2 \frac{d}{d\mu^2}\sum_{i,j} \mathcal{I}_{ij}(\bm{x})\mathfrak{S}_{\text{\tiny NS}}^{ij}=-a_s(\mu)\mathbb{H}_{\text{\tiny NS}}\sum_{i',j'} \mathcal{I}_{i'j'}(\bm{x})\mathfrak{S}_{\text{\tiny NS}}^{i'j'},
\end{eqnarray}
where $\mathbb{H}$ is integrated with $\mathcal{I}$. For values of $\bm{x}$ belonging to the grid $\bm{x}=\bm{x}(i,j)$ the equation simplifies
\begin{eqnarray}\label{ev-discrete}
\mu^2 \frac{d}{d\mu^2}\mathfrak{S}_{\text{\tiny NS}}^{ij}=-a_s(\mu)\sum_{i'j'}
{\mathbb{H}_{\text{\tiny NS}}}^{ij}_{i'j'}\mathfrak{S}_{\text{\tiny NS}}^{i'j'},
\end{eqnarray}
where
\begin{eqnarray}
{\mathbb{H}_{\text{\tiny NS}}}^{ij}_{i'j'}=\mathbb{H}_{\text{\tiny NS}}\mathcal{I}_{i'j'}(\bm{x}(i,j)).
\end{eqnarray}
The equation (\ref{ev-discrete}) is a system of ordinary linear differential equations which can be solved by standard methods (such as Runge-Kutta). The generalization for the singlet equation is straightforward. 

Thus, the most cumbersome part of the task is the construction of an appropriate grid and the interpolation function. They should provide reasonable numerical accuracy while remaining computationally efficient. We remind the reader that the goal is to create a code suitable for processing hundreds/thousands of values of physical observables, which set a limitation of seconds for a single evolution process.  The simplest solution would be equispaced grids in the $\bm{x}$ space, such as the grid implemented in ref.\cite{Pirnay:2013fra}. However, we found this choice unsatisfactory. The reason is singular/rapid behaviour of the solution in specific regions. We concluded that typically twist-3 PDFs have more rapid behavior around
\begin{itemize}
\item The central point $x_1=x_2=x_3=0$.
\item The edges of sectors $x_1=0$ or $x_2=0$ or $x_3=0$.
\end{itemize}
These sectors are of essential importance in practice. In particular, the central point corresponds to a ``small-x'' region of conventional PDFs, and thus is the main domain of high-energy experiments. The values of function at the edges corresponds to various projections needed for physical observables (for example, the Qiu-Sterman function \cite{Qiu:1991pp,Qiu:1991wg}). Consequently, one needs a denser grid around these point and lines in order to reach the desired precision for practical applications. For equispaced grids, where the density is homogeneous, one places an exceedingly large number of points in the center of the hexagon sectors, where the PDFs should typically be slowly-changing. Simultaneously, the size of 4-dimensional tensors $\mathbb{H}$ grows rapidly with the increase of the number of nodes, which, in turn, increases the cost of computation. Therefore, it is required to use grids that are denser at the center and diagonals of the hexagon support, and sparse in the remaining part. After several attempts we found a class of grids suitable for the present kernels and functions. They are described in this section.

\subsection{Grids}

\begin{figure}[!hb]
    \centering
    \includegraphics[width=0.55\textwidth]{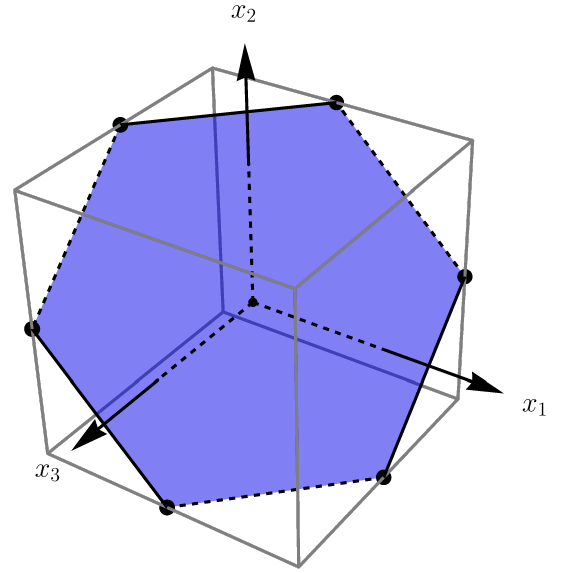}
    \caption{Graphical representation illustrating how the physical hexagonal support for the PDFs emerges by slicing the cube $|x_i|\le 1$ ($i=1,2,3$) with the plane defined by the momentum conservation condition $x_1+x_2+x_3$. }
    \label{fig_hexagon_from_cube}
\end{figure}

Due to the specific behavior of the evolution equations described in the previous section, it is clear that the natural notion of distance from the origin of a point is not given by the standard Euclidean norm, but rather the infinite norm. Specifically, we can measure the `radius' of a point as
\[\label{norm}
\lVert \bm{x} \rVert_{\infty} = \max(|x_1|,|x_2|,|x_3|).
\]
In this norm, the set of points equidistant from the origin forms a cube. The slice of the cube satisfying the momentum conservation condition $x_1+x_2+x_3=0$ forms a regular hexagon, which, if we fix the radius to be $1$, is exactly the physical boundary of the support of twist-3 PDFs. A graphical representation illustrating the emergence of the hexagon from the slicing of the cube is given in fig. \ref{fig_hexagon_from_cube}.

A general observation is that QCD evolution equations enhance the region close to the origin and suppress the region close to the physical boundary.
This suggests to use as the basic degrees of freedom to discretize the radius and angle on the $x_1+x_2+x_3=0$ surface, rather than the $x_i$ coordinates directly. 
In the given norm we can easily establish the notion of `angular-equispaced' points. 
To build the intuition, we observe that in the standard Euclidean norm (that suggests a natural notion of angle) the angle between pairs of vectors is the same if their projections on the unit circle are equispaced. Extending this rule to our infinite norm we find that the `angles' are equal if the points are equispace along the perimeter of the hexagon, that is the generalization of the `circle' for the norm (\ref{norm}).

This gives us a natural two dimensional space in terms of the radius and the angle in the infinite norm:
\[
(x_1,x_2,x_3)\equiv \bm{x} \to (r,\phi)
\]
with 
\[
\begin{split}
r(\bm{x}) &= \max(|x_1|, |x_2|, |x_3|) \\
\phi(\bm{x}) &= \begin{cases}
    \frac{x_2}{r},              & \text{ for } x_1>0,   x_2\ge0, x_3<0,  \\
    \ta 1- \frac{x_1}{r} \tc,   & \text{ for } x_1\le0, x_2>0,   x_3<0,  \\
    \ta 3- \frac{x_2}{r} \tc,   & \text{ for } x_1<0,   x_2>0,   x_3\ge0,\\
    \ta 3- \frac{x_2}{r} \tc,   & \text{ for } x_1<0,   x_2\le0, x_3>0,  \\
    \ta 4+ \frac{x_1}{r} \tc,   & \text{ for } x_1\ge0, x_2<0,   x_3>0,  \\
    \ta 6+ \frac{x_2}{r} \tc,   & \text{ for } x_1>0,   x_2<0,   x_3\le0,\\
\end{cases}
\end{split}
\]
such that $0<r\leqslant 1$ and $0\leqslant \phi <6$. The inverse map is
\[
\bm{x}(r,\phi) = r \begin{cases}
    (1-\phi,\phi,-1)       & 0\le \phi <1        \\
    (1-\phi,1,\phi-2)          & 1\le \phi <2        \\
    (-1,3-\phi,\phi-2)         & 2\le \phi <3        \\
    (-4+\phi,3-\phi,1)    & 3\le \phi <4        \\
    (\phi-4,-1,5-\phi)         & 4\le \phi <5        \\
    (1,\phi-6,5-\phi)       & 5\le \phi <6        \\
\end{cases}
\]
This 2D space is suited for the discretization of the evolution problem, meaning that we set $\bm{x}(i,j) \equiv \bm{x}(\phi_i, r_j)$ for some choice of the grid points $\phi_i, r_j$. 

To be clear, a grid is a map from the sets of integers $[0,1,\cdots 6n-1]\times [0,1,\cdots,M]$ to the intervals $[0,6)\times[0,1]$ such that 
\begin{eqnarray}\label{grid1}
(i,j) \to ( \phi_i, r_j).    
\end{eqnarray}
For the angular variable the total number of points is such that, for fix index $j$, each triangle that forms the hexagon has exactly $n$ points. 
To detail the features of the near-the-origin region, the radial grids should be chosen, in general, as non-linear functions. Due to the radial nature of evolution equation, it is simple to exclude the origin (or a small region near it) where typically there are no data and therefore no way to fix the distributions. Most grids in which the origin is excluded rely on some minimal value of $x$. 

A desirable aspect of the radial grid is a high density near the origin, and a not too sparse density near the physical boundary. As a default option that has these characteristics we use an hyperbolic cosine map (a similar map is successfully used in \texttt{artemide} code \cite{Scimemi:2019cmh, Moos:2023yfa}).
Specifically we use
\[
r_j = \qa \text{cosh}\ta \frac{j-M}{M c}  \tc\qc^{-\alpha}, \quad c^{-1} = -\text{acosh}\ta\frac{1}{x_{\text{\tiny min}}^{1/\alpha}}\tc
\label{hyp_j_to_r_map}
\]
where $c$ is the constant that fixes $r_0 = x_{\text{\tiny min}}$ and a typical good value for $\alpha$ is $3$. The inverse map is
\[
j = M\qa 1+c \  \text{acosh}\ta\frac{1}{r_j^{1/\alpha}}\tc\qc
\label{j_from_r}
\]

For the angular grid, the first and easiest option to consider is equispaced points. This is realized simply by choosing 
\[
\phi_i = \frac{i \text{$\:$mod$\:$} n}{n} + \qa \frac{i}{n}\qc,
\]
where the square brackets indicate integer division.
The inverse map is quite trivially
\[
i = n \phi_i.
\label{i_from_phi}
\]
 Other options that are available in the implementations are described in the appendix \ref{sec_app_different_grids}.

Generalizing the equations \eqref{j_from_r} and \eqref{i_from_phi} to continuous values of the radius and angle one obtains the two functions:
\begin{align}
\mathfrak{r}(\bm{x}) &= M\qa 1+c \  \text{acosh}\ta\frac{1}{(r(\bm{x}))^{1/\alpha}}\tc\qc \\
\mathfrak{f}(\bm{x}) &= n \phi(\bm{x}),
\end{align}
with the obvious properties 
\[
\mathfrak{f}(\bm{x}(i,j)) \equiv \phi_i, \quad \mathfrak{r}(\bm{x}(i,j)) \equiv r_j.
\]
In the following we refer to the space spanned by $\mathfrak{f},\mathfrak{r}$ as the grid-coordinate space. 

\subsection{Interpolation}

Having fixed a class of grids and the maps that transform from physical space to grid-coordinate space, the next step towards building the discretization of the evolution problem is to fix an interpolation function on the chosen grid. In the present work, we operate only with piece-wise linear interpolations, due to their numerical simplicity. This choice, due to the high locality of the interpolating functions, produces sparse evolution kernels. The trade-off of this choice is that a sizable number of points is required to reach good interpolation accuracy. The generalization to a higher-order interpolations seems to be an involved problem due to the non-Euclidean nature of the grid. The interpolation accuracy is the main precision limitation of our approach, and the implementation of a more accurate interpolation could be a good point for future investigations.

The aim of the interpolation is to write a (generically well-behaved) function of $\bm{x}$ as 
\[
f(\bm{x}) = \sum_{i=0}^{6n-1}\sum_{j=0}^M f_{ij}\mathcal{I}_{ij}(\bm{x}), \quad f_{ij}\equiv f(\bm{x}(i,j))
\label{inteprolation_generic}
\]
The weight functions $\mathcal{I}_{ij}(\bm{x}) \equiv \mathcal{I}_{ij}(\mathfrak{f}(\bm{x}),\mathfrak{r}(\bm{x}))$ are piece-wise linear functions in the \textit{transformed} coordinates $\mathfrak{f},\mathfrak{r}$, see eqn. \eqref{Iij_definition}. With our choice for $\mathfrak{r}(\bm{x})$ this implies that the weights $\mathcal{I}_{ij}(\bm{x})$ are non-linear functions of the physical momentum fractions $\bm{x}$.

\begin{figure}[!ht]
\centering
\begin{subfigure}{0.55\textwidth}
\begin{tikzpicture}
\tikzmath{\s = 1.5;}

\coordinate (ei) at (1.0*\s, 0.0*\s);
\coordinate (ej) at (0.0*\s, 1.0*\s);
\coordinate (FP) at ($-2*(ei)$);
\coordinate (FM) at ($+2*(ei)$);

\draw[white,fill=white] ($-4.2*(ei)-2.2*(ej)$) circle (5pt);
\draw[white,fill=white] ($+4.2*(ei)+2.8*(ej)$) circle (5pt);

\draw[-latex,thick,red] ($-4.0*(ei)$) -- ($4.0*(ei)$);
\draw[-latex,thick,red] ($-1.8*(ej)$) -- ($2.6*(ej)$);

\node[right] at ($2.6*(ej)$) {\Large $\mathfrak{r}$};
\node[above] at ($4.0*(ei)$) {\Large $\mathfrak{f}$};

\draw[thick] ($(FP) + 1.5*(ei) + 0.0*(ej)$) -- 
        ($(FP) + 1.5*(ei) - 1.5*(ej)$) -- 
        ($(FP) + 0.0*(ei) - 1.5*(ej)$) -- 
        ($(FP) - 1.5*(ei) + 0.0*(ej)$) -- 
        ($(FP) - 1.5*(ei) + 1.5*(ej)$) -- 
        ($(FP) + 0.0*(ei) + 1.5*(ej)$) --
        ($(FP) + 1.5*(ei) + 0.0*(ej)$);

\fill[red,opacity=0.25] ($(FP)$) -- ($(FP) + 1.5*(ei) - 1.5*(ej)$) -- ($(FP) + 0.0*(ei) - 1.5*(ej)$) -- ($(FP)$);

\draw[thick,dashed] ($(FP)$) -- ($(FP) + 1.5*(ei) + 0.0*(ej)$);
\draw[thick,dashed] ($(FP)$) -- ($(FP) + 1.5*(ei) - 1.5*(ej)$);
\draw[thick,dashed] ($(FP)$) -- ($(FP) + 0.0*(ei) - 1.5*(ej)$);
\draw[thick,dashed] ($(FP)$) -- ($(FP) - 1.5*(ei) + 0.0*(ej)$);
\draw[thick,dashed] ($(FP)$) -- ($(FP) - 1.5*(ei) + 1.5*(ej)$);
\draw[thick,dashed] ($(FP)$) -- ($(FP) + 0.0*(ei) + 1.5*(ej)$);

\draw[black, fill=black] ($(FP) + 1.5*(ei) + 0.0*(ej)$) circle (2pt);
\draw[black, fill=black] ($(FP) + 1.5*(ei) - 1.5*(ej)$) circle (2pt);
\draw[black, fill=black] ($(FP) + 0.0*(ei) - 1.5*(ej)$) circle (2pt);
\draw[black, fill=black] ($(FP) - 1.5*(ei) + 0.0*(ej)$) circle (2pt);
\draw[black, fill=black] ($(FP) - 1.5*(ei) + 1.5*(ej)$) circle (2pt);
\draw[black, fill=black] ($(FP) + 0.0*(ei) + 1.5*(ej)$) circle (2pt);

\node[right] at ($(FP)+(-0.075,0.22)$) {$\textcolor{blue}{(\mathfrak{f},\mathfrak{r})=(i,j)}$};
\node[left] at ($(FM)+(+0.075,0.22)$) {$\textcolor{blue}{(\mathfrak{f},\mathfrak{r})=(i,j)}$};

\node[above] at ($(FP) + 0.0*(ei) + 1.6*(ej)$) {\Large base of $\mathcal{I}^+_{ij}(\bm{x})$};
\node[above] at ($(FM) + 0.0*(ei) + 1.6*(ej)$) {\Large base of $\mathcal{I}^-_{ij}(\bm{x})$};

\draw[thick] ($(FM) + 1.5*(ei) + 0.0*(ej)$) -- 
        ($(FM) + 0.0*(ei) - 1.5*(ej)$) -- 
        ($(FM) - 1.5*(ei) - 1.5*(ej)$) -- 
        ($(FM) - 1.5*(ei) + 0.0*(ej)$) -- 
        ($(FM) + 0.0*(ei) + 1.5*(ej)$) -- 
        ($(FM) + 1.5*(ei) + 1.5*(ej)$) --
        ($(FM) + 1.5*(ei) + 0.0*(ej)$);
        
\draw[thick, dashed] ($(FM)$) -- ($(FM) + 1.5*(ei) + 0.0*(ej)$);
\draw[thick, dashed] ($(FM)$) -- ($(FM) + 0.0*(ei) - 1.5*(ej)$); 
\draw[thick, dashed] ($(FM)$) -- ($(FM) - 1.5*(ei) - 1.5*(ej)$); 
\draw[thick, dashed] ($(FM)$) -- ($(FM) - 1.5*(ei) + 0.0*(ej)$); 
\draw[thick, dashed] ($(FM)$) -- ($(FM) + 0.0*(ei) + 1.5*(ej)$); 
\draw[thick, dashed] ($(FM)$) -- ($(FM) + 1.5*(ei) + 1.5*(ej)$);

\draw[black, fill=black] ($(FM) + 1.5*(ei) + 0.0*(ej)$) circle (2pt);
\draw[black, fill=black] ($(FM) + 0.0*(ei) - 1.5*(ej)$) circle (2pt); 
\draw[black, fill=black] ($(FM) - 1.5*(ei) - 1.5*(ej)$) circle (2pt); 
\draw[black, fill=black] ($(FM) - 1.5*(ei) + 0.0*(ej)$) circle (2pt); 
\draw[black, fill=black] ($(FM) + 0.0*(ei) + 1.5*(ej)$) circle (2pt); 
\draw[black, fill=black] ($(FM) + 1.5*(ei) + 1.5*(ej)$) circle (2pt);
        
\draw[black,fill=blue] ($(FP)$) circle (3pt);
\draw[black,fill=blue] ($(FM)$) circle (3pt);



\end{tikzpicture}
\caption{}
\label{fig_two_grid}
\end{subfigure}
\begin{subfigure}{0.35\textwidth}
\includegraphics[width=\textwidth]{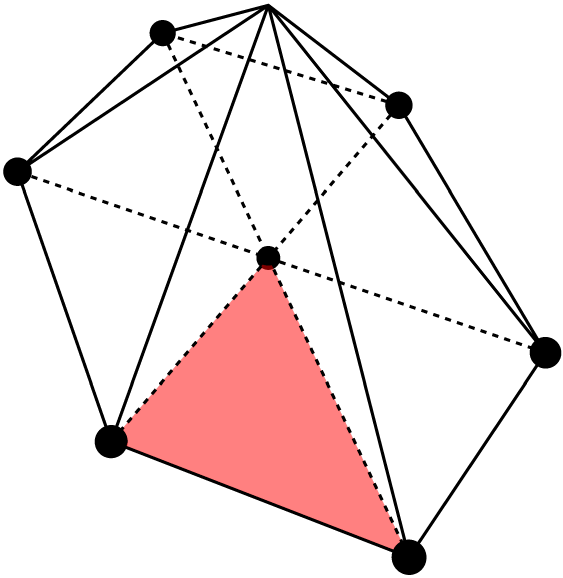}
\caption{}
\label{fig_HexagonsInPhysicalSpace}
\end{subfigure}
\caption{Panel (a) shows the subdivision in triangles of the basis of the two weight functions $\mathcal{I}^\pm_{ij}$. Panel (b) shows the 3D representation of the weight function $\mathcal{I}^+_{ij}$. The vertex of the pyramid has height equal to $1$. The red shaded triangle is used to guide with the orientation.}
\label{figure_interpolants}
\end{figure}
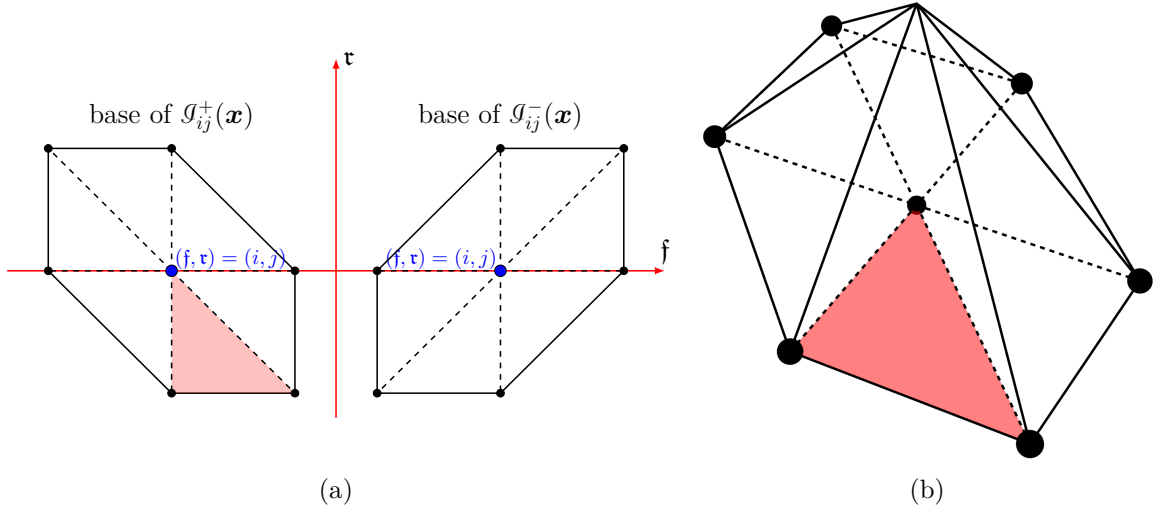

To build a piece-wise linear weight function $\mathcal{I}_{ij}$ in the $\mathfrak{f},\mathfrak{r}$ variables it is sufficient to consider grid points with $i\pm1, j\pm1$ (understanding the periodic nature of the index $i$ and the restrictions in the index $j$). With these nine points (eight on the edge and the center point $i,j$), there is still freedom on how to construct the linear interpolation. Our choice is to define two different weight functions: 
\[
\label{Iij_definition}
\mathcal{I}^\pm_{ij}(\bm{x}) = \text{max} \bigg[0,1-\text{max}\Big(|\mathfrak{f}(\bm{x})-i|\ ,\ |\mathfrak{r}(\bm{x})-j|\ ,\ |\mp \mathfrak{f}(\bm{x})-\mathfrak{r}(\bm{x})\pm i+j|\Big) \bigg]
\]
each of which uses a subset of six out of the eight neighbors of $(i,j)$. 
These functions are pyramids with a non-regular hexagonal base. In fig. \ref{figure_interpolants} we show the graphical representations of these functions. In panel (a) we show the basis of the pyramids and in panel (b) the three-dimensional graph for $\mathcal{I}^+_{ij}$ (the graph for $\mathcal{I}^-_{ij}$ is similar).
For instance, the piece of the $\mathcal{I}_{ij}^+$ weight function that has as base the triangle formed by the three grid points $(i,j), (i,j-1), (i+1,j-1)$ has the following explicit expression:
\[
\mathcal{I}^+_{ij}(\bm{x})  = \mathfrak{r}(\bm{x}) -j+1 \quad \text{ if } \mathfrak{f}(\bm{x})\ge i, \ \mathfrak{r}(\bm{x}) > j-1, \ \mathfrak{f}(\bm{x})+\mathfrak{r}(\bm{x}) < i+j
\]
The base triangle is shaded in red in fig. \ref{figure_interpolants}. From this expression it is clear that if the point $(\mathfrak{f}(\bm{x}),\mathfrak{r}(\bm{x}))$ remains inside the triangle the weight function is linear.

Using either solely $\mathcal{I}^+_{ij}$ or solely $\mathcal{I}^-_{ij}$ disturbs the symmetries of the task with respect to reflections, resulting into undesired artifacts in the solution. Thus, to preserve the natural symmetries and to reduce inaccuracies, we utilize the composition of both, and construct our interpolation as
\[
f(\bm{x}) = \sum_{i=0}^{6n-1} \sum_{j=0}^M f_{ij} \frac{\mathcal{I}^+_{ij}(\bm{x})+\mathcal{I}^-_{ij}(\bm{x})}{2}.
\]

Finally, with the interpolation for a generic function defined, we can discretize the evolution kernels. The generic expression is
\[
H_{i'j'}^{ij} = \frac{H_{i'j'}^{+,ij} + H_{i'j'}^{-,ij}}{2} = \frac{1}{2}\ta \mathbb{H}\mathcal{I}^+_{i'j'}(\bm{x}(i,j))+ \mathbb{H}\mathcal{I}^-_{i'j'}(\bm{x}(i,j))\tc.
\]
The evolution equation for, e.g. the non-singlet, becomes
\[
\mu^2\frac{\partial}{\partial\mu^2}\mathfrak{S}^{ij}_{\text{\tiny NS}_i} = -a_s(\mu^2) \sum_{i'=0}^{6n-1}\sum_{j'=0}^M \ta H_{\text{\tiny NS}}\tc_{i'j'}^{ij} \mathfrak{S}^{i'j'}_{\text{\tiny NS}_i}
\]
and similarly for the chiral-odd and the singlet cases.

We have found that a grid of size $60\times15$ is sufficient for an average relative interpolation precision of $10^{-2}$, while $120\times25$ provides precision $10^{-3}$. The highest uncertainty of reconstruction is concentrated at $r(\bm{x})\to 1$, if we exclude the outer part of hexagon ($r(\bm{x})<0.9$) the average interpolation precision became $\sim 10^{-4}$ (Note that in the physical case, the absolute values for the PDFs at $r(\bm{x})\to 1$ are small, and turn to zero at the boundary). These tests have been performed using the model in eq. \eqref{eq_pdf_model}. Comparing with other performed tests, we have concluded that the interpolation is the main source of numerical uncertainty in our code. In what follows we use grids $120\times25$ for tests, and consider it as a default choice. 

To determine the discretized kernels one has to carry out the defying integrals given in appendix \ref{sec_app_evolutionkernels}. The numerical evaluation is straightforward, once a few delicate points have been addressed. To highlight these points, we consider a part of $\widehat{\mathcal{H}}_{12}$ kernel. For the sake of shortness we denote $x_{1,2,3}^{ij}\equiv x_{1,2,3}(i,j)$ and focus on the action on the $\mathcal{I}^+$ function, since the same considerations apply to $\mathcal{I}^-$. The kernel is
\[
\widehat{\mathcal{H}}_{12} \mathcal{I}^+_{i'j'}(\bm{x}^{ij}) = \int_{-\infty}^{\infty} dv \frac{x^{ij}_1\Theta(x^{ij}_1,-v)}{v(v-x^{ij}_1)} \qa \mathcal{I}^+_{i'j'}(\bm{x}^{ij})-\mathcal{I}^+_{i'j'}(x_1^{ij}-v,x_2^{ij}+v,x_3^{ij})\qc
\]
The first term in the square brackets performs the subtraction that makes the integral well-defined at $v=0$. The subtraction term has an integration range over the whole real line, while for other terms the integration range is limited by the support of $\mathcal{I}^+_{i'j'}$.   

It can be directly checked that our choice for the weight functions satisfies
\[
\mathcal{I}^+_{i'j'}(\bm{x}^{ij}) = \delta_{ii'}\delta_{jj'}\mathcal{I}^+_{ij}(\bm{x}^{ij}),
\]
which naturally leads to a decomposition of the kernel in a diagonal, proportional to $\delta_{ii'}\delta_{jj'}$ and a non-diagonal parts. Let us start with the latter, for which the subtraction constant is absent:
\[
\int_{-\infty}^{\infty} dv \frac{x^{ij}_1\Theta(x^{ij}_1,-v)}{v(x^{ij}_1-v)} \mathcal{I}^+_{i'j'}(x_1^{ij}-v,x_2^{ij}+v,x_3^{ij}).
\]
This integral could be computed numerically as is. However, to increase the accuracy and speed of the evaluation, it is preferable to identify the limits of the integration as they are restricted by the support of the interpolant $\mathcal{I}^+_{i'j'}$. Due to the non-linear nature of the grid, determining the exact support can be cumbersome. Instead, we can easily find a (sometimes) slightly larger range $[v_{\text{\tiny min}},v_{\text{\tiny max}}]$ which is guaranteed to contain the support in $v$ and is of comparable size.
Given these values we solve the simple system of equations
\[
\begin{split}
& x_1^{\text{\tiny min}}\le x_1^{ij}-v \le x_1^{\text{\tiny max}}, \quad x_2^{\text{\tiny min}}\le x_2^{ij}+v \le x_2^{\text{\tiny max}}, x_3^{\text{\tiny min}}\le x_3 \le x_3^{\text{\tiny max}} \Rightarrow \\
& v_{\text{\tiny min}} = \text{max}\ta x_1^{ij}-x_1^{\text{\tiny max}}, x_2^{\text{\tiny min}} - x_2^{ij}\tc, \quad v_{\text{\tiny max}} = \text{min}\ta x_1^{ij}-x_1^{\text{\tiny min}}, x_2^{\text{\tiny max}} - x_2^{ij}\tc.
\end{split}
\]
We stress that the minimal and maximal values are computed from the $i'j'$ point, not the $i,j$ one. Consequently, the non-diagonal part of the kernel is
\[
\begin{cases}
\displaystyle \int_{v_{\text{\tiny min}}}^{v_{\text{\tiny max}}} dv \frac{x^{ij}_1\Theta(x^{ij}_1,-v)}{v(x^{ij}_1-v)} \mathcal{I}^+_{i'j'}(x_1^{ij}-v,x_2^{ij}+v,x_3^{ij})
 & \text{ if }  x_3^{ij}\in[x_3^{\text{\tiny min}}, x_3^{\text{\tiny max}}], v_{\text{\tiny min}}<v_{\text{\tiny max}} \\
 0 & \text{ otherwise }
\end{cases}
\]
The determination of $v_{\text{\tiny min,max}}$ is obviously the same for the diagonal part of the kernel, that can now be decomposed as
\begin{align}
&\int_{v_{\text{\tiny min}}}^{v_{\text{\tiny max}}} dv \frac{x^{ij}_1\Theta(x^{ij}_1,-v)}{v(v-x^{ij}_1)} \qa \mathcal{I}^+_{ij}(\bm{x}^{ij})-\mathcal{I}^+_{ij}(x_1^{ij}-v,x_2^{ij}+v,x_3^{ij})\qc \notag\\
& + \mathcal{I}^+_{ij}(\bm{x}^{ij}) \qa \int_{-\infty}^{v_{\text{\tiny min}}} dv+\int_{v_{\text{\tiny max}}}^{\infty} dv \qc \frac{x^{ij}_1\Theta(x^{ij}_1,-v)}{v(v-x^{ij}_1)}.
 \end{align}

The integrals in the second line can be computed analytically and give
\[
\mathcal{I}^+_{ij}(\bm{x}^{ij})  \qa \theta(x^{ij}_1)\log \ta1-\frac{x^{ij}_1}{v_{\text{\tiny min}}}\tc + \theta(-x^{ij}_1)\log \ta 1-\frac{x^{ij}_1}{v_{\text{\tiny max}}}\tc\qc,
\]
where we incidentally notice that both logs vanish for $x^{ij}_1=0$.
Therefore, the final expression for the kernel is
\begin{align}
\widehat{\mathcal{H}}_{12}\mathcal{I}^+_{i'j'}(\bm{x}^{ij}) &= \ta\widehat{H}_{12}\tc_{i'j'}^{+,ij} = \delta_{i\neq i' \text{ or } j\neq j'}\int_{v^{i'j'}_{\text{\tiny min}}}^{v^{i'j'}_{\text{\tiny max}}} dv \frac{x^{ij}_1\Theta(x^{ij}_1,-v)}{v(x^{ij}_1-v)} \mathcal{I}^+_{i'j'}(x_1^{ij}-v,x_2^{ij}+v,x_3^{ij})  \notag\\
& + \delta_{ii'}\delta_{jj'} \mathcal{I}^+_{ij}(\bm{x}^{ij})  \qa \theta(x^{ij}_1)\log \ta1-\frac{x^{ij}_1}{v^{ij}_{\text{\tiny min}}}\tc + \theta(-x^{ij}_1)\log \ta 1-\frac{x^{ij}_1}{v^{ij}_{\text{\tiny max}}}\tc\qc \notag\\
& +  \delta_{ii'}\delta_{jj'}\int_{v^{ij}_{\text{\tiny min}}}^{v^{ij}_{\text{\tiny max}}} dv \frac{x^{ij}_1\Theta(x^{ij}_1,-v)}{v(v-x^{ij}_1)} \qa \mathcal{I}^+_{ij}(\bm{x}^{ij})-\mathcal{I}^+_{ij}(x_1^{ij}-v,x_2^{ij}+v,x_3^{ij})\qc\notag
\end{align}
where we added the superscripts to $v_{\text{\tiny min,max}}$ to avoid any confusion on which are the values of $x_{\text{\tiny min,max}}$ used for their computation.

This procedure is to be carried out in identical fashion for the other part of $\widehat{\mathcal{H}}_{12}$ and for all other kernels. Note that the rank-four tables of evolution kernels are sparse (with typical density of non-zero entries of $\sim 10^{-3} - 10^{-4}$), therefore, we use dedicated sparse algebra algorithms for storage and multiplication of these kernels. The kernels are computed during the initialization stage of the codes and internally stored. Initialization times vary based on how many computational cores are available, but typically using $\sim$ 10 cores the initialization can be expected to finish in about $1\sim10$ minutes.  

The codes require the initial values of PDFs, initial and final scales, and the function $\alpha_s(\mu)$. Then it discretizes the input functions, and combine them into C-parity-definite functions with flavor-singlet and non-singlet combinations. Then each flavor combination is evolved using a fourth-order Runge-Kutta algorithm. Results are combined into the definite-flavor combinations and delivered to the user.

\section{Example evaluation}
\label{sec_results}

\begin{figure}[!ht]
    \centering
    \includegraphics[width=0.85\textwidth]{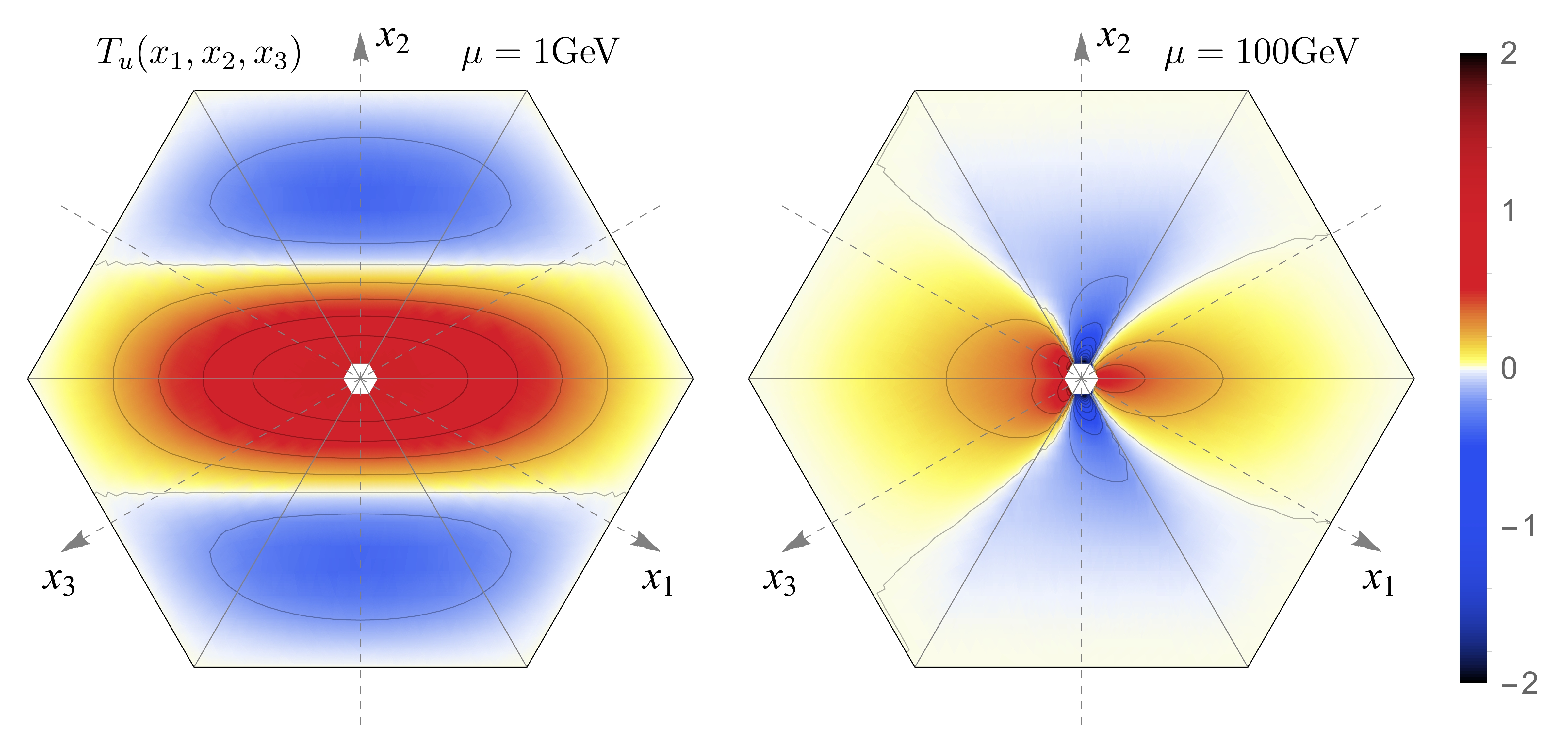}
    \caption{Evolution of the flavor-up distribution $T_u$ of eqn.\eqref{def:T} from $1$ GeV to $100$ GeV. The boundary conditions are specified in eqn.(\ref{eq_pdf_model}).}
    \label{fig_Tu_1_to_100}
\end{figure}

\subsection{Example of evolution}
We have run and compared our codes in various circumstances, including stress tests (rapid functions, very dense grids, very small values of $x_{\text{min}}$, etc), and found a solid agreement between two independent implementations (both presented in the sec.\ref{sec_code_interface}).  The main source of numerical disagreement (of order of $\sim 10^{-7}$) is the differences in the implementation of the integral kernels, which is of the same order as the aimed numerical precision. If identical kernels are used, the agreement between implementations is of the order of $10^{-16}$.

Generally, we have observed that the evolution effects are large in the bulk of the support of the function, and result into significant changes of the functional form, especially in the vicinity of the origin. In this section, we demonstrate a simple example of results of evaluation.

One of the problems for the test is that a preferred physical form of twist-3 PDFs is unknown, since yet there are no phenomenological fits (the problem which is aimed by this project). Thus, for the demonstration, we have chosen a rather simple form of boundary condition:
\begin{align}
\mathcal{h}(x_{1,2,3}) &= (1-x_1^2)(1-x_2^2)(1-x_3^2) \notag\\
T_u(x_{1,2,3}) &= \cos(4x_2)\mathcal{h}(x_{1,2,3})\quad
T_d(x_{1,2,3}) = \ta 2 - \cos(3\pi \mathcal{h}(x_{1,2,3}))\tc \mathcal{h}(x_{1,2,3}) \notag\\ 
\Delta T_u(x_{1,2,3}) &= \mathcal{h}(x_{1,2,3}) \ta \sin(\pi x_2) + 4(x_1^2 - x_3^2)\tc\notag\\
\label{eq_pdf_model}
\Delta T_d(x_{1,2,3}) & = 2\frac{\sin(\pi x_2)}{\|\bm{x}\|_{\infty}}\ta 1-\cos(\mathcal{h}(x_{1,2,3}))\tc\\
T_s(x_{1,2,3}) & = -\frac{3}{10}T_d(x_{1,2,3})\quad 
\Delta T_s(x_{1,2,3}) = -\frac{3}{10}\Delta T_d(x_{1,2,3})\notag\\
T^+(x_{1,2,3}) &= \mathcal{h}(x_{1,2,3})\|\bm{x}\|_{\infty} \sin(x_1-x_3)\quad 
T^-(x_{1,2,3}) = \mathcal{h}(x_{1,2,3}) \|\bm{x}\|_{\infty} \cos(x_1-x_3)\notag
\end{align}
These functions obey all required symmetries and turn to zero at $||\mathbf{x}||_{\infty}=1$. They are rather smooth at the origin, which could be (in)consistent with the physics.

\begin{figure}
    \centering
    \includegraphics[width=0.45\textwidth]{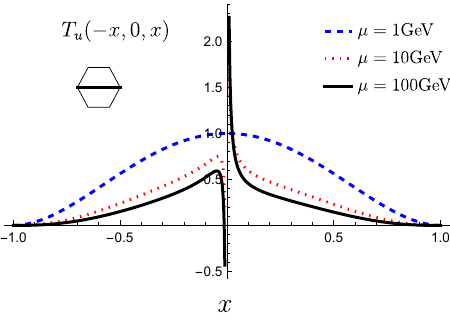}\ 
    \includegraphics[width=0.45\textwidth]{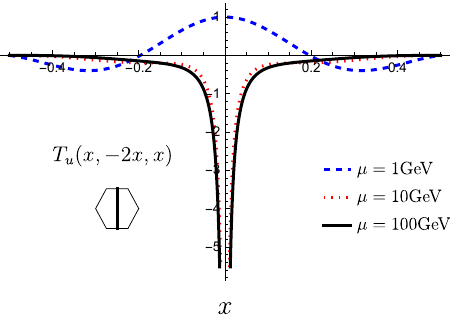}\\ 
    \includegraphics[width=0.45\textwidth]{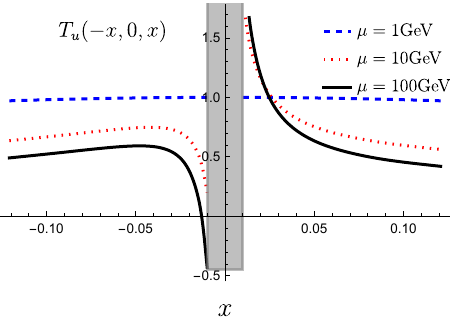}\phantom{\ 
    \includegraphics[width=0.45\textwidth]{Figures/T_up_1to10_N.pdf}}
    \caption{Comparison of two slices of $T_u$ when evolved from $1$ GeV to $100$ GeV: on the left $T_u(-x,0,x)$ with at the top the full range and at the bottom a zoomed-in version near the origin; on the right $T_u(x,-2x,x)$. The gray band displays the excluded region in $x$ due to the imposed constraint $x_{\text{min}}=10^{-2}$. The small hexagons show the path of the slice on the two-dimensional support.}
    \label{fig_Tu_1_to_100_slices}
\end{figure}

To gauge the effect of evolution we present the $T_u$ distribution. In Fig.\ref{fig_Tu_1_to_100} we show the boundary condition at a scale $\mu = 1$ GeV on the left and the result of its evolution to $\mu = 100$ GeV on the right. As one can see, the effect of evolution is particularly pronounced also for moderate values of $\|\bm{x}\|_{\infty}$, even if most of the effect is concentrated for small values of $\|\bm{x}\|_{\infty}$, where the shape of distribution is modified. To have a clearer picture of the effects of evolution, in fig.\ref{fig_Tu_1_to_100_slices} we display two mono-dimensional sections of $T_u$. We show $T_u(-x,0,x)$ (which is the Qiu-Sterman function \cite{Qiu:1991pp,Qiu:1991wg}) on the left panel, and $T(x,-2x,x)$ on the right panel, which is the function along the line orthogonal to the Qiu-Sterman case (i.e. along the $x_2$ axis in Fig.\ref{fig_Tu_1_to_100}). We observe that the evolution has a strong effect in $\|\bm{x}\|_{\infty}\lesssim 10^{-1}$ region. This region is  important in practice: for example, almost half of the data for the SIDIS transverse single-spin asymmetry, which is described by the distribution $T$, is measured at $\|\bm{x}\|_{\infty}\sim 0.1 - 0.05$ (see e.g. reviews of data in refs.\cite{Bury:2021sue, Horstmann:2022xkk}), at scales from $\sim 1 $GeV (for example \cite{JeffersonLabHallA:2011ayy}) till $\sim 90$GeV (for example \cite{STAR:2015vmv}). Specifically for the Qiu-Stermann $T_u(-x,0,x)$ the effect of the evolution is very pronounced, which has important consequences for many applications. From fig.\ref{fig_Tu_1_to_100_slices} it is also evident that it is essential to have a grid that manages the low $\|\bm{x}\|_\infty$ region accurately in order to achieve  good accuracy for the evolution.

\newcommand{\bjorn}{\cite{Pirnay:2013fra} }
\subsection{Comparison with the existing code of \texorpdfstring{\cite{Pirnay:2013fra}}{}}

We have also compared against the code presented with the preprint \bjorn\footnote{We are grateful to V. Braun and A. Manashov for sharing with us a few known bugs in the implementation of $\mathbb{H}_{gg}^\pm$ kernels in \bjorn. These bugs were also reported in \cite{Hatta:2019csj}.}. In ref.\bjorn the discretization is realised with an equispaced grid in the variables $x_1,x_2$. Albeit being a natural starting point, that we also employed in the first iteration of our implementation, it suffers from very low accuracy near $r(\bm{x})\sim0$ point. We remind the reader, that this region is essential for the measurements at high energies. The problem can be partially mitigated by increasing the number of points in the grid. However, due to the scaling of the kernels as the fourth power of the number of points, this solution is not practical. Also, the kernel discretization in \bjorn was performed using the Simpson integration rule on the grid points, which greatly reduces the accuracy of the kernels with respect to the Gauss-Kronrod adaptive quadrature that we employ. Lastly, in ref.\bjorn the system of differential equations is solved by the direct exponentiation of the kernels, whereas our implementation is the 4th order Runge-Kutta algorithm, which has better precision. 

Comparing the results of the test runs of our codes and the one of ref.\bjorn, we have found them generally compatible (especially for $r\gtrsim 0.2$, where both implementations have similar grid densities). The agreement however is still only in the $10\%$ range. For smaller values of $r$ the differences are larger. We have not found any bug or improper behavior in our implementations, which were independently written and cross-compared in the end. Therefore, we can only conclude that the approach of \cite{Pirnay:2013fra}, albeit being in principle correct, suffers from large numerical inaccuracies.

\section{Code interface}
\label{sec_code_interface}
The two codes ({\tt C} and {\tt FORTRAN}) are to be separately compiled and are completely independent.
We strove for two different implementations with the same underlying logic in such a way that a cross-check between the implementations gives automatically a check of the implementations themselves.
With the code is shipped a pre-computed set of kernels for $n=20$ and $M=25$, i.e. $120\times25$ total points for the solution.

The codes are available at \cite{honeycomb} for the \texttt{C} implementation and at \cite{snowflake} for the \texttt{FORTRAN} implementation.  In the two following subsections we provide the necessary details to install and run the code. Please also refer to the tests files that are provided with each implementation for any further details.

\subsection{C}
\newcommand{\ccode}[1]{\lstinline[language=C]{#1}}
\newcommand{\shcode}[1]{\lstinline[language=sh]{#1}}

\label{sec_code_interface_sub_C}
The code, dubbed `honeycomb' from the hexagonal support for the PDFs, is available at \cite{honeycomb}.
In the following, we present a description of how to interface with the library. More details and example evaluation can be found in the test files which are available in the \shcode{tests/src} directory. 

To operate with the library the user must specify the input parameters and initialize the interface. 
\begin{itemize}
\item  The input parameters must be initialized first. This can be done directly in the code, or by reading a configuration card calling
\begin{lstlisting}[language=C]
input_parameters_t *in_par = 
    init_input_parameters(conf_path, to_save)
\end{lstlisting}
where {\tt conf\_path} is the path to the configuration card and {\tt to\_save} is a boolean to control whether the configuration used should also be outputted to file.

\item To initialize the interface itself the user must call
\begin{lstlisting}[language=C]
evolution_interface_t *ei = 
    init_evolution_interface(as, as_par, 
                             ev, in_par, v_l, ei, prev);
\end{lstlisting}
where: 
\begin{itemize}
\item[{\it i)}] {\tt as} and {\tt as\_par} are a function to compute $\alpha_s/4\pi$ and its parameters respectively.
\item[{\it ii)}] {\tt ev} is a pointer to an \ccode{evolution_functions_t} struct, which is a bundle of functions pointers to the relevant functions for the computation of the boundary conditions.
\item [{\it iii)}] {\tt in\_par} is a pointer to the \ccode{input_parameters_t} struct. 
\item [{\it iv)}] {\tt v\_l} is a verbose level, which can be \ccode{PL_NONE, PL_ESSENTIAL, PL_ALL}.
\item [{\it v)}] \ccode{ei} is a pointer to an \ccode{evolution_interface_t} struct.
\item [{\it vi)}] \ccode{prev} is a boolean that signals whether the previously stored state in the interface should be used as a starting point.
\end{itemize}
\end{itemize}

If the pointer to the evolution interface, which is passed as the argument, is set to NULL, then the interface is forced to perform complete initialization. Otherwise, if a valid pointer is given, the interface attempts to re-use a part of the information that is stored. If the interface is to be initialized, the boundary conditions are re-computed from the input functions no matter the value of the boolean passed as the last parameter to the initialization. 

The review of the configuration options is given in appendix \ref{app_config_C}. If any is left empty or given in the wrong format, the program will resort to default configuration for the problematic option. The order in which the parameters are given in the configuration card is irrelevant.

The \ccode{evolution_functions_t} fields follow the naming convention ${\tt Tf} \leftarrow T_f $, ${\tt DTf} \leftarrow \Delta T_f $ for the PDF of flavor $f$ and ${\tt TFp} \leftarrow T^+_{3F} $, ${\tt TFm} \leftarrow T^-_{3F} $ for the two gluon distributions. The parameter pointer is \ccode{p_PDF}. 
\textbf{Important: the model must satisfy the physical symmetry relations} (see sec. \ref{sec_qgq_pdfs} and sec. \ref{sec_ggg_pdfs}), \textbf{otherwise an incorrect result is obtained.} During the initialization procedure of the evolution interface there is no automatic detection of the violation of the symmetries. To test the chosen model for violation of the symmetries one can use the \ccode{check_*_symmetry} functions, with ${\tt *}={\tt T, DT, TFp, TFm}$. These functions return either true (symmetry respected) or false (symmetry broken). If a point breaks the symmetries, its value is logged. The check is performed on a sample of $10^4$ random points inside the physical hexagon. If the $x_{\text{\tiny min}}$ value is invalid, the program is aborted.

To execute the evolution one can use either calls
\begin{lstlisting}[language=C]
saved_solution_t *solOut = execute_evolution(ei);
saved_solution_t **solOut_v = execute_evolution_w_steps(ei);
\end{lstlisting}
where the first one only saves the result at the final scale, whereas the second one saves the result at each step of the evolution.
To export the solution to disk
one can call
\begin{lstlisting}[language=C]
save_model(solOut, prefix, basis);
\end{lstlisting}
where \ccode{prefix} is a string that is prefixed to the filename and \ccode{basis} specifies the basis in which the solution is to be saved, valid options are the physical basis \ccode{BASIS_PHYSICAL}, the definite-C-parity basis \ccode{BASIS_DEFINITE_C_PAR} or both \ccode{BASIS_BOTH}. A previously computed model can be loaded via the  \ccode{load_model} call.

Finally, to release the memory associated with the evolution interface, saved solution and input parameters, the user should make the calls
\begin{lstlisting}
free_evolution_interface(&ei);
free_saved_solution(&solOut);
free_input_parameters(&in_par);
\end{lstlisting}

\subsection{FORTRAN}
\label{sec_code_interface_sub_FORTRAN}

The FORTRAN code has the name \textit{Snowflake}, due to similarity of the twist-three support region with a snowflake (see fig.\ref{fig_HexagonsInPhysicalSpace}). The code is available at \cite{snowflake}. It is stand-alone and requires compiler FORTRAN 95 or newer. To compile the code one needs to update the compilator call in \textit{makefile} (by default \texttt{f95}), and call \texttt{make}. The correctness of compilation can be checked by calling \texttt{make test}, which runs an elementary code and tests the proper linking between modules. The resulting object-files are stored in \texttt{/obj}. A simple single-files code can be compiled with \textit{Snowflake} by calling \texttt{make program TARGET=...}, where dots are the path to the source code. An example of such a simple code can be found in \texttt{/prog/EXAMPLE.f90}.

The main module is \texttt{snowflake}, which contains several public functions and subroutines. The first subroutine to be called is
\begin{center}
\texttt{call Snowflake\_Initialize(file,path)}
\end{center}
Here, \texttt{file} is the name of INI-file, which contains all required setup information, and \texttt{path} is the optional argument with the path to the \texttt{file} (if \texttt{path} is absent the INI-file is searched in the directory of the program). The INI-file has specific format. The default configuration is presented in \texttt{snowflake.ini} in the root directory. The description of the INI-file is given in appendix \ref{app_config_FORTRAN}.

The initialization procedure allocates memory, and loads them from the storage, path to which is given in the INI-file. If the loading of kernels is switched off, the kernels will be computed. The latter can take significant time about 10-15 min for default configuration on a typical desktop. Kernels can be computed using the program \texttt{prog/saveKernels.f90}.

The computation of evolution is made by
\begin{center}
\texttt{call ComputeEvolution(mu0,mu1,alpha,G1,U1,D1,S1,C1,B1,
G2,U2,D2,S2,C2,B2,inputQ,inputG)}
\end{center}
where \texttt{mu0} and  \texttt{mu1} are initial and final scales in GeV. \texttt{alpha} is a single-variable function which represents $\alpha_s(\mu)$. It has interface \texttt{real*8:: alpha, mu}. The arguments \texttt{G1,U1,D1,S1,C1,B1,G2,U2,D2,S2,C2,B2} are optional, and represents two-variable functions \texttt{f(x1,x2)} with interfaces \texttt{real*8::f,x1,x2}. These functions are the boundary values for the evolved distribution at \texttt{mu0}. They are interpreted with accordance to optional parameters \texttt{inputQ} (accepts values \texttt{'T'}, \texttt{'S'} or \texttt{'C'}) and \texttt{inputG} (accepts values \texttt{'T'} or \texttt{'C'}). The interpretation rules are
\begin{center}
\begin{tabular}{c|l | c|l|l}
function    &  flag & interpretation & definition & symmetry
\\\hline
\texttt{U1} & \texttt{inputQ='T'} & $T_u(x_1,x_2)$ & (\ref{def:T},\ref{frakS->T},\ref{S->T}) & (\ref{symm:T-quark})
\\\hline
\texttt{U2} & \texttt{inputQ='T'} & $\Delta T_u(x_1,x_2)$ & (\ref{def:deltaT},\ref{frakS->T},\ref{S->T}) & (\ref{symm:T-quark})
\\\hline
\texttt{U1} & \texttt{inputQ='S'} & $S^+_u(x_1,x_2)$ & (\ref{S+<-frakS<-T}) & (\ref{symm:S-quark})
\\\hline
\texttt{U2} & \texttt{inputQ='S'} & $S^-_u(x_1,x_2)$ & (\ref{S-<-frakS<-T}) & (\ref{symm:S-quark})
\\\hline
\texttt{U1} & \texttt{inputQ='C'} & $\mathfrak{S}^+_u(x_1,x_2)$ & (\ref{frakS<-T<-S}) & (\ref{symm:frakS-quark})
\\\hline
\texttt{U2} & \texttt{inputQ='C'} & $\mathfrak{S}^-_u(x_1,x_2)$ & (\ref{frakS<-T<-S}) & (\ref{symm:frakS-quark})
\\\hline
\multicolumn{5}{p{10.5cm}}{\texttt{D1,D2,S1,S2,C1,C2,B1,B2} represent the boundary values of for $d$, $s$, $c$ and $b$ flavors correspondingly with the same interpretation rules.}
\\\hline
\texttt{G1} & \texttt{inputG='T'} & $T_{3F}^{+}(x_1,x_2)$ & (\ref{braun_to_us_T3F},\ref{T=F})
& (\ref{symm:T-gluon})
\\\hline
\texttt{G2} & \texttt{inputG='T'} & $\Delta T_{3F}^-(x_1,x_2)$ & (\ref{braun_to_us_T3F},\ref{T=F})
& (\ref{symm:T-gluon})
\\\hline
\texttt{G1} & \texttt{inputG='C'} & $\mathfrak{F}^+_u(x_1,x_2)$ & (\ref{F<-T(gluon)})
& (\ref{symm:F-gluon})
\\\hline
\texttt{G2} & \texttt{inputG='C'} & $\mathfrak{F}^-_u(x_1,x_2)$ & (\ref{F<-T(gluon)})
& (\ref{symm:F-gluon})
\end{tabular}    
\end{center}
If the argument is absent the corresponding boundary value is zero. The default values for flags are \texttt{'C'}. The call of procedure evolves the boundary values to \texttt{mu1} and stores in the memory.

\textbf{Important:} The boundary-value functions must satisfy the physical symmetry relations. Otherwise, the result of computation is incorrect. There is no internal check for consistency.

The value of the evolved function at $(x_1,x_2,-x_1-x_2)$ can be obtained by call of the functions
\begin{center}
\texttt{real*8 GetPDF(x1,x2,f,outputT)}
\end{center}
where \texttt{x1} and  \texttt{x2} are \texttt{real*8} variables $x_1$ and $x_2$. The integer \texttt{f} defines the type of the function with accordance to optional flag \texttt{outputT} (default value is \texttt{'C'}). The type is selected with respect to the following table
\begin{center}
\begin{tabular}{l|c | c| c| c| c| c| c| c| c| c}
\texttt{f} & n & -n & n & -n & n & -n & 0 or 10 & -10 & 0 or 10 & -10
\\ \hline
\texttt{outputT=} & \texttt{'T'} & \texttt{'T'} & \texttt{'S'} & \texttt{'S'} 
& \texttt{'C'} & \texttt{'C'} & \texttt{'T'} & \texttt{'T'} & \texttt{'C'} & \texttt{'C'}
\\ \hline
return  & $T_q$& $\Delta T_q$ & $S^+_q$& $S^-_q$ & $\mathfrak{S}^+_q$& $\mathfrak{S}^-_q$ 
& $T_{3F}^+$ & $T_{3F}^-$ & $\mathfrak{F}^+$ & $\mathfrak{F}^-$
\\  \hline
\multicolumn{11}{c}{
n=1,2,3,4,5 and corresponds to quark flavors $d$, $u$, $s$, $c$, $b$, correspondingly.}
\end{tabular}
\end{center}

To compute the evolution of functions $H$ or $E$ (\ref{def:HE}) call the subroutine
\begin{center}
\texttt{ComputeEvolutionChiralOdd(mu0,mu1,alpha,U1,D1,S1,C1,B1)}    
\end{center}
where \texttt{mu0} and  \texttt{mu1} are initial and final scales in GeV. \texttt{alpha} is a single-variable function which represents $\alpha_s(\mu)$. The boundary conditions \texttt{U1,D1,S1,C1,B1} are two-variable functions, which are interpreted as zero if absent. These functions are always interpreted as $H$ or $E$. Since $H$ and $E$ obey the same evolution equation, they are distinguished only by symmetries (\ref{symm:HE}). The result at $(x_1,x_2)$ is invoked by
\begin{center}
\texttt{real*8 GetPDFChiralOdd(x1,x2,f)}   
\end{center}
where \texttt{x1} and  \texttt{x2} are \texttt{real*8} variables $x_1$ and $x_2$. The integer argument \texttt{f}=1,2,3,4,5 and corresponds to the flavors $d$, $u$, $s$, $c$, $b$, correspondingly. The type of output is dictated by the input, if the boundary values were symmetric (i.e. satisfy $H(x_{123})=H(-x_{321})$) the evolved function is $H$, and similar for function $E$. Note that the linearity of evolution equation implies that functions $H$ and $E$ can be evolved simultaneously, applying evolution to a general function those (anti-)symmetric part corresponds to $H$ ($E$).

\section{Conclusions}
\label{sec_conclusions}

This work presents an open-source, general purpose code for the evolution of twist-3 parton distribution functions (PDFs). The code is fully tested and could be easily  integrated into other codes of a high-energy library. This is a prerequisite for a systematic analysis of structure functions that involves twist-3 PDFs. Various experimental measurements are sensitive to different projections of twist-3 PDFs, which are functions of two kinematic variables. Simultaneously, the total number of kinds of PDFs is not large: there are four types of quark and two types of gluon PDFs. Thus, a global analysis of a variety of measurements can restrict these functions along various lines, while the evolution and loop corrections will relate these values to each other. Altogether, it opens the possibility for the first experimental determination of parton interference effects in the nucleon. The code-base presented in this work provides one with tools to perform this analysis, giving the solutions for evolution, storage, discretization, and manipulations with twist-3 PDFs.

The paper also includes a review, in sec. \ref{sec_definitions}, of the fundamental definitions and parametrizations of twist-3 PDFs. This translation dictionary can be used to relate various studies, since there is no standard choice of parametrization. This happens because there is no single definition for twist-3 objects that highlights all the relevant features. For instance, the evolution equations simplify dramatically in the definite C-parity basis, which is, however, quite unwieldy for perturbative computations of coefficient functions (see e.g.\cite{Braun:2021aon, Rein:2022odl}). In the same section we also provided an intuitive description of the geometry of the evolution equations, which is quite different compared to the standard twist-2 evolution, due to the two-dimensional support of the twist-3 PDFs.

The two-dimensional nature of the twist-3 PDFs requires adaptation and modification of the standard techniques that are well established for solving twist-2 evolution. In sec. \ref{sec_discretizaitons} we provided ample explanation to the natural radial-angular coordinate system that we used to define a discretization of the problem. Such a coordinate system allows one to impose a higher density of grid points towards the physical origin, which is essential for future phenomenological applications. Moreover, the union of the coordinate system and choice of interpolation functions ensure that the radial ordering of the kernels, described in the sec. \ref{sec_definitions} is maintained in the discretized problem. 

We decided to implement the evolution problem in two separate and independent codes. One is written in the {\tt FORTRAN} language (and called \texttt{snowflake}), the other in the {\tt C} language (and called \texttt{honeycomb}). This double implementation not only helps lowering the barrier to the usage of the codes in different projects, but also allowed us to cross-check implementations. The latter is important since the only another evolution code for twist-3 PDFs is available \cite{Pirnay:2013fra}, and does not provide sufficient accuracy. Our implementation can be easily incorporated in larger existing projects and it is independent of any external library (except from {\tt libomp}). The codes are available at \cite{honeycomb} for the \texttt{C} implementation and at \cite{snowflake} for the \texttt{FORTRAN} implementation.

\section*{Acknowledgements}
We thank V. Braun and A. Manashov for the useful comments on the properties of twist-3 PDFs and their evolution. SR is supported by the Deutsche Forschungsgemeinschaft (DFG, German Research Foundation) – grant number 409651613 (Research Unit FOR 2926), subproject 430915355 and grant number 491245950. A.V. is funded by the \textit{Atracci\'on de Talento Investigador} program of the Comunidad de Madrid (Spain) No. 2020-T1/TIC-20204 and \textit{Europa Excelencia} EUR2023-143460, MCIN/AEI/10.13039/501100011033/,  from Spanish Ministerio de Ciencias y Innovaci\'on.

\appendix
\section{Different grids}
\label{sec_app_different_grids}

In this appendix, we present all types of grids that are currently implemented in the code.  The grid along the radius and the angle can be specified independently, and thus we present them independently. To select the type of the grid in the C-code, one must specify fields \texttt{radial\_grid\_type}, and \texttt{angular\_grid\_type}  in the configuration file accordingly. To select the type of the grid in the Fortran-code, one must modify the compiler variables \texttt{GRIDTYPE\_R} and \texttt{GRIDTYPE\_PHI} at the beginning of the \texttt{HexGrid.f90} file, and recompile the library. 

The default option for the radial grid is given by the hyperbolic cosine map in  eqn.\eqref{hyp_j_to_r_map}.

The second option for the radial grid is the `log-exp' type of grid, where
\[
\label{eq_logexp_grid}
r_j = \exp\qa -\log(x_{\text{\tiny min}}) \frac{j-M}{M}\qc, \quad j\in[0,M],
\]
being $x_{\text{\tiny min}}$ the minimal allowed value for the radius. 
This grid has the obvious feature that $r_0 = x_{\text{\tiny min}}$. 
The inverse transformation is 
\[
j = M\qa 1-\frac{\log(r(\bm{x}))}{\log(x_{\text{\tiny min}})}\qc.
\]

The third option of the radial grid is the `root-power' type of grid, where we define
\[
r_j = \ta\frac{\frac{j}{M} + c}{1+c}\tc^{\alpha}.
\]
For the $j=M$ point grid reaches the maximum value $r_{M} = 1$, automatically. The minimum value is defined by parameter $c$, which is fixed such that $r_0=x_{\text{\tiny min}}$.  We find
\[
\label{eq_rootpower_grid}
c = \frac{x_{\text{\tiny min}}^{1/\alpha}}{1-x_{\text{\tiny min}}^{1/\alpha}}.
\]
The inverse transformation is
\[
j = M\qa (1+c)\ta r(\bm{x})\tc^{1/\alpha}-c\qc.
\]
The parameter $\alpha$ controls the distribution points. For $\alpha>1$, the points are denser at $x\to x_{\text{min}}$. The case for $\alpha=1$ represents the radially-equispaced case.

One can see that the log-exp grid is very dense near the origin, and somewhat sparse near the physical boundary. The root-power grid is less dense near the origin and denser near the physical boundary (the exact behavior depends on the choice of $\alpha$, where $0<\alpha<1$ makes the grid sparse near the origin and should be avoided in practice). Ideally, we would keep the same density near the origin as the log-exp grid and have a grid with a comparable density as the root-power one with $\alpha\sim2,3$ near the physical boundary. This can be achieved by modifying the log-exp grid and adding a second log.
Specifically, we can take
\[
\label{eq_iml_grid}
r_j = 2\frac{\exp\qa \frac{j-M}{Mc}\qc}{1+\exp\qa \frac{j-M}{Mc}\qc},
\]
which inverts to
\[
j = M\qa 1 + c \log\ta\frac{r_j}{2-r_j}\tc\qc, \quad c^{-1} = -\log\ta\frac{x_{\text{\tiny min}}}{2-x_{\text{\tiny min}}}\tc.
\]
This grid exhibits the low-$x$ behavior of the log-exp grid and the large-$x$ behavior of the root-power grid.

\begin{figure}[!ht]
    \centering
    \includegraphics[width=0.55\textwidth]{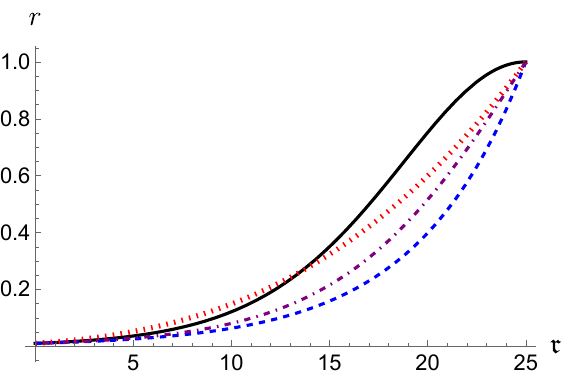}
    \caption{The solid black line is the hyperbolic cosine map of  eqn.\eqref{hyp_j_to_r_map}, the blue dashed line is the log-exp map of  eqn.\eqref{eq_logexp_grid}, the red dotted line is the root-power grid of  eqn. \eqref{eq_rootpower_grid}, and the purple dot-dashed line is the double-log map of  eqn. \eqref{eq_iml_grid}.  All grid are showed using $x_{\text{min}}=10^{-2}$, $M=25$ and for the hyperbolic cosine and root-power grids the $\alpha$ exponent is fixed to $3$.}
    \label{fig_grid_comp}
\end{figure}

To help the comparison between the different grids we show in fig. \ref{fig_grid_comp} the (continuous) map from $\mathfrak{r}$ to $r$ for each specific grid choice. The flatter a line in the plot is, the denser is the corresponding grid. It is evident how all the choices have comparable densities near the origin, but only the hyperbolic cosine map has also a satisfactory density near the physical border (which corresponds to the right-most part of the abscissa).

\begin{figure}[!hb]
\centering
\begin{subfigure}{0.45\textwidth}
\centering
\includegraphics[width=\textwidth]{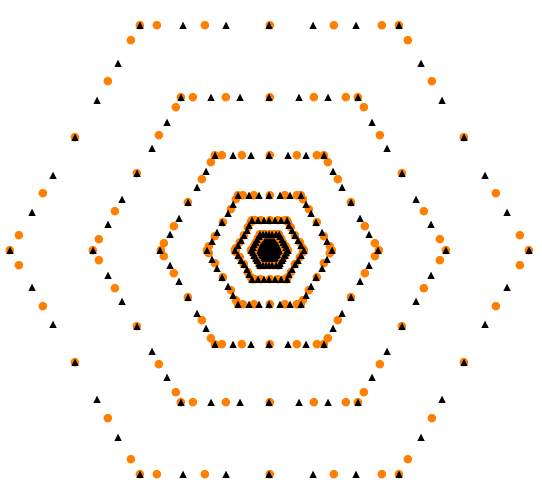}
\caption{}
\label{fig_GridLINvsCOS_IML}
\end{subfigure}
\begin{subfigure}{0.45\textwidth}
\centering
\includegraphics[width=\textwidth]{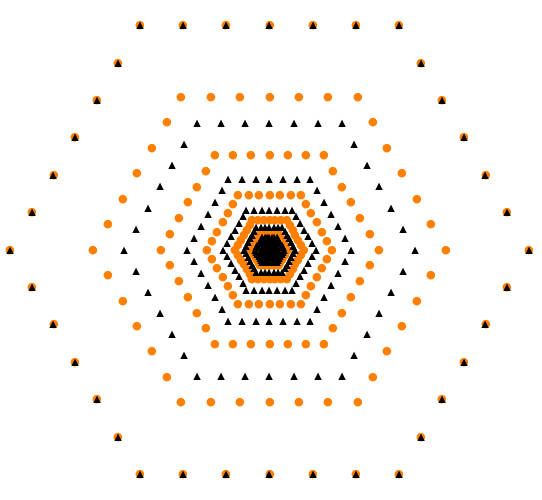}
\caption{}
\label{fig_GridIMLvsLOG_LIN}
\end{subfigure}
\caption{Panel (a) is the graphical representation of the grid with cosine angular scaling (orange dots) and linear angular scaling (black triangles) both using the double-log radial grid. Panel (b) is the graphical representation of the double-log grid (orange dots) and the single-log grid (black triangles) both using the linear angular grid}
\label{fig_GridComp}
\end{figure}

For the angular grids, an alternative to the equispaced points is a map that increases the density of points towards the lines $x_{1,2,3}=0$. For instance, one can use a cosine map:
\[
\phi_i = \frac{1}{2}-\frac{1}{2}\cos\ta\pi \frac{i \text{$\:$mod$\:$} n}{n}\tc + \qa \frac{i}{n}\qc
\]
with an inverse that is a bit more cumbersome than the simple linear map:
\[
i=\qa \phi(\bm x)\qc n+\frac{n}{\pi}\text{arccos}    \qa 1- 2 \text{$\:$mod$\:$}(\phi(\bm x),1) \qc
\]
We show the two-dimensional comparison for different choices of grids in fig.\ref{fig_GridComp}. 

\section{Evolution kernels}
\label{sec_app_evolutionkernels}
In this appendix, we collect the expressions for the evolution kernels in momentum space. For their presentation, we use the notation
\[
\Theta(a,b) \equiv \theta(a)\theta(b) - \theta(-a)\theta(-b)
\]
where 
\[
\theta(a) = \begin{cases}
    1 & a>0 \\ 0 & a\le 0
\end{cases}
\] 
The condition $\theta(0)=0$ secures the numerical stability of the implementation but introduces the necessity of treating the points $x_i=0$ (where $x_i$ is one of the generic momentum fractions) carefully. To isolate singular contributions we introduce the function 
\[
\delta_{x_i,0} = \begin{cases}
    1 & \text{ if } x_i=0, \\
    0 & \text{ otherwise.}
\end{cases}
\]
We observe two different types of behaviors at $x_i=0$. 
\begin{itemize}
\item First, the kernels are continuous and well-defined for $x_i=0$, but a direct numerical implementation would suffer from inaccuracies introduced by the cancellation of large contributions of different signs. This happens, for instance, in the $\mathcal{H}^+$ kernels. To provide accurate and reliable results we isolate the problematic point and compute the kernel separately. The isolation of the problematic point generates a structure of the form $\delta_{x_i,0}(\cdots) + \delta_{x_i\neq 0}(\cdots)$.
\item Second, the final expression for the kernel is ambiguous and the point $x_i=0$ cannot be straightforwardly isolated. This happens solely for the pieces of the $\widehat{\mathcal{H}}$ kernels when one of the $x_i$ is a gluon momentum fraction and for the exchange kernel $\mathcal{H}_{23}^{e,(1)}$. To reliably treat these points, we have re-derived the momentum-space expressions starting from the unambiguous position-space expression \cite{Braun:2009mi}, explicitly considering $x_i=0$ case. The isolation of the problematic point generates a structure of the form $\delta_{x_i,0}(\cdots) + (\cdots)$, where there is no $\delta_{x_i\neq0}$ contribution.
\end{itemize}

The transformation from the expressions in ref.\cite{Braun:2009mi} to the one presented in the following is achieved by using the variable transformation $v_i = x_l -v$ $v_j = x_k + v$, where for the most of cases $x_l=x_i$, $x_k=x_j$, except for the kernels marked with `$-$' and `$e$', where $x_l=x_j$ and $x_k = x_i$. The latter identification also clarifies why such kernels are referred to as `exchange kernels'. After the transformation, the three-variable $\Theta$-functions of \cite{Braun:2009mi} can be reduced to two-variable $\Theta$-functions with a little algebra.

\subsection{Non-singlet kernels}
For the leading-color contributions we have
\begin{align}
\widehat{\mathcal{H}}_{12}\mathfrak{S} &=\delta_{x_2,0}\mathfrak{S}(x_1,0,x_3)\\
& + \int dv \Bigg[ -\frac{x_1\Theta(x_1,-v)}{v(x_1-v)} \ta\mathfrak{S}(x_1,x_2,x_3)- \mathfrak{S}(x_1-v,x_2+v,x_3)\tc\notag \\
& \quad \quad+ \frac{x_2\Theta(x_2,v)}{v(v+x_2)} \ta\mathfrak{S}(x_1,x_2,x_3)- \frac{x_2}{v+x_2}\mathfrak{S}(x_1-v,x_2+v,x_3)\tc \Bigg] \notag\\
\widehat{\mathcal{H}}_{23}\mathfrak{S} &=\delta_{x_2,0}\mathfrak{S}(x_1,0,x_3)\\
& + \int dv \Bigg[ -\frac{x_3\Theta(x_3,-v)}{v(x_3-v)} \ta\mathfrak{S}(x_1,x_2,x_3)- \mathfrak{S}(x_1,x_2+v,x_3-v)\tc\notag\\
& \quad \quad+ \frac{x_2\Theta(x_2,v)}{v(v+x_2)} \ta\mathfrak{S}(x_1,x_2,x_3)- \frac{x_2}{v+x_2}\mathfrak{S}(x_1,x_2+v,x_3-v)\tc \Bigg] \notag\\
\mathcal{H}^+_{12} \mathfrak{S} &= \delta_{x_3,0}  \int  dv   \Theta(x_1,-v) \frac{v (v-2 x_1)}{2 (x_1-v)^3}  \mathfrak{S}(x_1-v,-x_1+v,0)\\
& +
\delta_{x_3\neq0} \int dv \Bigg[ \Theta(x_2,v)\frac{x_2^2(x_2-x_3+v)}{2(x_2+v)^2x_3^2} \notag \\&  
\phantom{+
\delta_{x_3\neq0} \int dv }
+ \Theta(x_1,-v)  \frac{x_1(x_2-x_3)}{2(x_1-v)x_3^2}\Bigg] \mathfrak{S}(x_1-v,x_2+v,x_3)\notag 
\end{align}
For the color-suppressed contributions we have
\begin{align}
\widehat{\mathcal{H}}_{13}\mathfrak{S} &= \int dv  \ta \frac{x_3\Theta(x_3,v)}{v(x_3+v)}-\frac{x_1\Theta(x_1,-v)}{v(x_1-v)}  \tc \\& \notag \qquad\qquad \times \ta\mathfrak{S}(x_1,x_2,x_3)- \mathfrak{S}(x_1-v,x_2,x_3+v)\tc\\
\mathcal{H}^+_{13}\mathfrak{S} &= \delta_{x_2,0}\int  dv \Theta(-x_1,v)\frac{v}{(v-i)^2} \mathfrak{S}(x_1-v,0,-x_1+v)\\
& +\delta_{x_2\neq0}\int dv \Bigg[ \frac{x_1}{x_2(v-x_1)}\Theta(x_1,-v) \notag \\
& \phantom{+\delta_{x_2\neq0}\int dv} +\frac{-x_3}{x_2(x_3+v)}\Theta(x_3,v) \Bigg]  \mathfrak{S}(x_1-v,x_2,x_3+v)\notag \\
\mathcal{H}^-_{12}\mathfrak{S} &=\delta_{x_3,0} \int dv \Theta(x_1,-v)\frac{v^2}{2(x_1-v)^3} \mathfrak{S}(-x_1+v,x_1-v,0) \\
& \notag +\delta_{x_3\neq0}\int dv \Bigg[ \Theta(x_1,-v)\frac{x_1(2x_2(x_1-v)-x_1(x_2+v))}{2(x_1-v)^2x_3^2} \\
& \notag \phantom{+\delta_{x_3\neq0}\int dv}  + \Theta(x_2,v)\frac{x_2^2}{2(v+x_2)x_3^2}\Bigg] \mathfrak{S}(x_2+v,x_1-v,x_3)\\
\mathcal{H}^{e,(1)}_{23}P_{23}\mathfrak{S} &= \delta_{x_3,0}\mathfrak{S}(x_1,0,x_2) +\int dv \frac{x_3}{(x_3-v)^2}\Theta(x_3,-v)\mathfrak{S}(x_1,x_3-v,x_2+v)
\end{align}
\subsection{Chiral-odd kernels}
Compared to the non-singlet chiral even case, only two new kernels appear
\begin{align}
\mathcal{H}^+_{23}  H &=  \delta_{x_10} \int dv   \Theta(x_3,-v) \frac{v (v-2 x_3)}{2 (x_3-v)^3}  H(0,-x_3+v,x_3-v)\\
& +\delta_{x_1\neq 0}\int dv \Bigg[ \Theta(x_2,v)\frac{x_2^2(x_2-x_1+v)}{2(x_2+v)^2x_1^2} \notag \\
& \qquad \qquad + \Theta(x_3,-v)  \frac{x_3(x_2-x_1)}{2(x_3-v)x_1^2}\Bigg] H(x_1,x_2+v,x_3-v) \notag
\end{align}
\begin{align}
\mathcal{H}^-_{23}  H & =  \int dv \Bigg[ \Theta(x_3,-v)\frac{x_3(2x_2(x_3-v)-x_3(x_2+v))}{2(x_3-v)^2x_1^2} \\
& \notag \qquad \qquad + \Theta(x_2,v)\frac{x_2^2}{2(v+x_2)x_1^2}\Bigg] H(x_1,x_3-v,x_2+v) \notag\\
&= \delta_{x_10}  \int dv \Theta(x_3,-v) \Bigg[-\frac{v^2}{2 \left(v-x_3\right)^3}\Bigg]H(0,x_3-v,-x_3+v)\\
& +\delta_{x_1\neq 0}\int dv \Bigg[ \Theta(x_3,-v)\frac{x_3(2x_2(x_3-v)-x_3(x_2+v))}{2(x_3-v)^2x_1^2} \notag \\
& \notag \qquad \qquad \qquad + \Theta(x_2,v)\frac{x_2^2}{2(v+x_2)x_1^2} \Bigg] H(x_1,x_3-v,x_2+v)\notag
\end{align}

\subsection{Singlet kernels: diagonal terms}
For the diagonal quark-quark case we only need one extra kernel compared to the non-singlet, which reads
\[
\mathcal{H}^d_{13}\mathfrak{S} =  -\Theta(x_1,x_3)\frac{x_1x_3}{x_2^3} \int dv \mathfrak{S}(x_1-v,x_2,x_3+v)
\]
For the diagonal gluon-gluon case we need similar kernels as for the non-singlet case, but modified to accommodate the different conformal spin of the particle
\begin{align}
\widehat{\mathcal{H}}_{12}  \mathfrak{S} &=  \delta_{x_10} \mathfrak{S}(0,x_2,x_3)+ \delta_{x_20} \mathfrak{S}(x_1,0,x_3) \\
&  +  \int dv \Bigg[ - \frac{x_1\Theta(x_1,-v)}{v(x_1 - v)} \ta\mathfrak{S}(x_1,x_2,x_3)- \frac{x_1}{x_1 - v}\mathfrak{S}(x_1 -v,x_2 + v,x_3)\tc \notag\\
& \qquad\qquad+ \frac{x_2\Theta(x_2,v)}{v(x_2 + v)} \ta\mathfrak{S}(x_1,x_2,x_3)- \frac{x_2}{x_2+v}\mathfrak{S}(x_1-v,x_2+v,x_3)\tc \Bigg]\notag
\end{align}

Quite evidently we have that
\begin{align}
\widehat{\mathcal{H}}_{23}  \mathfrak{S} &= \delta_{x_20} \mathfrak{S}(x_1,0,x_3) + \delta_{x_30} \mathfrak{S}(x_1,x_2,0) \\
&+ \int dv \Bigg[ - \frac{x_2\Theta(x_2,-v)}{v(x_2 - v)} \ta\mathfrak{S}(x_1,x_2,x_3)- \frac{x_2}{x_2 - v}\mathfrak{S}(x_1,x_2-v,,x_3+v)\tc \notag \\
& \qquad\qquad + \frac{x_3\Theta(x_3,v)}{v(x_3 + v)} \ta\mathfrak{S}(x_1,x_2,x_3)- \frac{x_3}{x_3+v}\mathfrak{S}(x_1,x_2-v,x_3+v)\tc \Bigg]\notag
\end{align}
and
\begin{align}
\widehat{\mathcal{H}}_{31}  \mathfrak{S} &= \delta_{x_10} \mathfrak{S}(0,x_2,x_3) + \delta_{x_30} \mathfrak{S}(x_1,x_2,0) \\
&+ \int dv \Bigg[ - \frac{x_3\Theta(x_3,-v)}{v(x_3 - v)} \ta\mathfrak{S}(x_1,x_2,x_3)- \frac{x_3}{x_3 - v}\mathfrak{S}(x_1+v,x_2,,x_3-v)\tc \notag \\
& \qquad\qquad+ \frac{x_1\Theta(x_1,v)}{v(x_1 + v)} \ta\mathfrak{S}(x_1,x_2,x_3)- \frac{x_1}{x_1+v}\mathfrak{S}(x_1+v,x_2,x_3-v)\tc \Bigg]\notag
\end{align}
For the ``$+$''-kernels we have
\begin{align}
\mathcal{H}^+_{12}  \mathfrak{S} &= \delta_{x_3 0}  \int dv \Theta(x_1,-v) \qa-\frac{v \left(-6 v x_1+6 x_1^2+v^2\right)}{6 \left(v-x_1\right)^4}\qc \mathfrak{S}(x_1-v,-x_1+v,0) \\
& + \delta_{x_3 \neq 0} \int dv \Bigg[ \Theta(x_1,-v) \ta \frac{x_1^2 \left(-v \left(x_1+3 x_2\right)+8 x_1 x_2+3 x_1^2+3 x_2^2\right)}{6 \left(x_1+x_2\right)^3 \left(v-x_1\right)^2}\tc \notag\\
& \qquad \qquad+ \Theta(x_2,v) \ta\frac{x_2^2 \left(x_2 \left(8 x_1+v\right)+3 x_1 \left(x_1+v\right)+3 x_2^2\right)}{6 \left(x_1+x_2\right)^3 \left(x_2+v\right)^2}\tc \Bigg] \notag \\& \qquad \qquad \qquad \times\mathfrak{S}(x_1-v,x_2+v,x_3)\notag \\
\widetilde{\mathcal{H}}^+_{12}  \mathfrak{S} &=  \delta_{x_30}\int dv \Theta(x_1,-v) \qa-\frac{v^3}{6 \left(v-x_1\right)^4}\qc \mathfrak{S}(x_1-v,-x_1+v,0) \\
& + \delta_{x_3\neq 0} \int dv \Bigg[ \Theta(x_1,-v) \ta -\frac{x_1^2 \left(v \left(x_1+3 x_2\right)-2 x_1 x_2\right)}{6 \left(x_1+x_2\right)^3 \left(v-x_1\right)^2}\tc \notag\\
& \qquad\qquad+ \Theta(x_2,v) \ta \frac{x_2^2 \left(v \left(3 x_1+x_2\right)+2 x_1 x_2\right)}{6 \left(x_1+x_2\right)^3 \left(x_2+v\right)^2}\tc \Bigg] \mathfrak{S}(x_1-v,x_2+v,x_3)\notag\\
\mathcal{H}^+_{13}  \mathfrak{S} &= \delta_{x_2 0}  \int dv \Theta(x_1,-v) \qa-\frac{v \left(-6 v x_1+6 x_1^2+v^2\right)}{6 \left(v-x_1\right)^4}\qc \mathfrak{S}(x_1-v,0,-x_1+v) \\
& + \delta_{x_2 \neq 0} \int dv \Bigg[ \Theta(x_1,-v) \ta \frac{x_1^2 \left(-v \left(x_1+3 x_3\right)+8 x_1 x_3+3 x_1^2+3 x_3^2\right)}{6 \left(x_1+x_3\right)^3 \left(v-x_1\right)^2}\tc \notag\\
& \qquad \qquad+ \Theta(x_3,v) \ta\frac{x_3^2 \left(x_3 \left(8 x_1+v\right)+3 x_1 \left(x_1+v\right)+3 x_3^2\right)}{6 \left(x_1+x_3\right)^3 \left(x_3+v\right)^2}\tc \Bigg] \notag \\& \qquad \qquad \qquad \times\mathfrak{S}(x_1-v,x_2,x_3+v)\notag
\\
\widetilde{\mathcal{H}}^+_{13}  \mathfrak{S} &=  \delta_{x_20}\int dv \Theta(x_1,-v) \qa-\frac{v^3}{6 \left(v-x_1\right)^4}\qc \mathfrak{S}(x_1-v,0,-x_1+v) \\
& + \delta_{x_2\neq 0} \int dv \Bigg[ \Theta(x_1,-v) \ta -\frac{x_1^2 \left(v \left(x_1+3 x_3\right)-2 x_1 x_3\right)}{6 \left(x_1+x_3\right)^3 \left(v-x_1\right)^2}\tc \notag\\
& \qquad\qquad+ \Theta(x_3,v) \ta \frac{x_3^2 \left(v \left(3 x_1+x_3\right)+2 x_1 x_3\right)}{6 \left(x_1+x_3\right)^3 \left(x_3+v\right)^2}\tc \Bigg] \mathfrak{S}(x_1-v,x_2,x_3+v)
\notag
\end{align}

For the ``$-$''-kernels we have the inversion as for the non-singlet case:
\begin{align}
\mathcal{H}_{12}^-  \mathfrak{S} &= \delta_{x_30} \Theta(x_1,-v) \int dv \Bigg[-\frac{v^3}{6 \left(v-x_1\right)^4}\Bigg]\mathfrak{S}(-x_1+v,x_1-v,0)\\
& +\delta_{x_3\neq 0}\int dv \Bigg[ \ta -\frac{x_1^2 \left(v \left(x_1+3 x_2\right)-2 x_1 x_2\right)}{6 \left(x_1+x_2\right)^3 \left(v-x_1\right)^2}\tc \Theta(x_1,-v)\notag  \\
& \qquad\qquad+ \ta \frac{x_2^2 \left(v \left(3 x_1+x_2\right)+2 x_1 x_2\right)}{6 \left(x_1+x_2\right)^3 \left(x_2+v\right)^2}\tc \Theta(x_2,v) \Bigg] \mathfrak{S}(x_2+v,x_1-v,x_3)\notag \\
\mathcal{H}_{13}^-  \mathfrak{S} &= \delta_{x_20} \Theta(x_1,-v) \int dv \Bigg[-\frac{v^3}{6 \left(v-x_1\right)^4}\Bigg]\mathfrak{S}(-x_1+v,0,x_1-v)\\
& +\delta_{x_2\neq 0}\int dv \Bigg[ \ta -\frac{x_1^2 \left(v \left(x_1+3 x_3\right)-2 x_1 x_3\right)}{6 \left(x_1+x_3\right)^3 \left(v-x_1\right)^2}\tc \Theta(x_1,-v)\notag  \\
& \qquad\qquad+ \ta \frac{x_3^2 \left(v \left(3 x_1+x_3\right)+2 x_1 x_3\right)}{6 \left(x_1+x_3\right)^3 \left(x_3+v\right)^2}\tc \Theta(x_3,v) \Bigg] \mathfrak{S}(x_3+v,x_2,x_1-v)\notag 
\end{align}

\subsection{Singlet kernels: quark-gluon terms}
For the two kernels $\mathcal{V}^\pm$ we have
\begin{align}
\mathcal{V}_{13}^+  \mathfrak{F} &= \delta_{x_20} \Theta(x_1,-v)\int dv \qa \frac{x_1^2}{\left(v-x_1\right)^4}\qc \mathfrak{F}(x_1-v,0,-x_1+v) \\
& + \delta_{x_2\neq 0}\int dv \Bigg[ \Theta(x_1,-v) \ta -\frac{x_1 x_3 \left(3 x_1+x_3-2 v\right)}{\left(x_1+x_3\right)^3 \left(v-x_1\right)^2}\tc  \notag\\
& \qquad \qquad+ \Theta(x_3,v) \ta \frac{x_1 x_3 \left(x_1+3 x_3+2 v\right)}{\left(x_1+x_3\right)^3 \left(x_3+v\right)^2}\tc  \Bigg] \mathfrak{F}(x_1-v,x_2,x_3+v) \notag\\
\mathcal{V}_{13}^-  \mathfrak{F} &= \delta_{x_20} \Theta(x_1,-v) \int dv \qa \frac{v^2}{\left(v-x_1\right)^4}\qc\mathfrak{F}(-x_1+v,0,x_1-v)\\
& + \delta_{x_2\neq0}\int dv \Bigg[ \Theta(x_1,-v)\ta\frac{x_1 \left(-x_1 x_3+x_1^2+2 v x_3\right)}{\left(x_1+x_3\right)^3 \left(v-x_1\right)^2}\tc   \notag\\
& \qquad \qquad + \Theta(x_3,v)\ta \frac{x_3 \left(x_3 \left(x_1-x_3\right)+2 v x_1\right)}{\left(x_1+x_3\right)^3 \left(x_3+v\right)^2}\tc \Bigg] \mathfrak{F}(x_3+v,x_2,x_1-v) \notag
\end{align}

\subsection{Singlet kernels: gluon-quark terms}
Here we give expressions for the un-permuted combinations. The action of the permutation operator $P_{23}$ exchanges $x_2\leftrightarrow x_3$.
we have
\begin{align}
\mathcal{W}^+   \mathfrak{S} &=   -\frac{1}{2}\int dv \qa \Theta(x_1,-v)  - \Theta(x_3,v)\qc \mathfrak{S}(x_1-v,x_2,x_3+v) \\
\mathcal{W}^-   \mathfrak{S} &=   \delta_{x_20} \int dv \Theta(x_1,-v)\qa \frac{v^2}{\left(v-x_1\right)^2}\qc  \mathfrak{S}(-x_1+v,0,x_1-v) \\
& + \delta_{x_2\neq 0} \int dv \Bigg[ \Theta(x_1,-v) \frac{2x_1^2+\left(x_1+x_3\right) \left(v-x_1\right)}{2\left(x_1+x_3\right) \left(v-x_1\right)}  \notag \\
& \qquad \qquad +  \Theta(x_3,v) \frac{2x_3^2-\left(x_1+x_3\right) \left(x_3+v\right)}{2\left(x_1+x_3\right) \left(x_3+v\right)}\Bigg] \mathfrak{S}(x_3+v,x_2,x_1-v) \notag\\
\Delta\mathcal{W}  \mathfrak{S} &= \delta_{x_3\neq 0}\int dv \qa \Theta(x_3,v) - \Theta(x_1,x_3) x_1^2 \frac{3x_3 + x_1}{(x_1+x_3)^3}\qc\mathfrak{S}(x_1-v,x_2,x_3+v)
\\\notag &-\delta_{x_30} \int dv\Theta(x_1,-v) \mathfrak{S}(x_1-v,x_2,v)
\end{align}

\section{Configuration file for {\tt C} implementation}
\label{app_config_C}
The following options can be specified:
\begin{center}
\begin{longtable}{c|p{2.5cm}|p{5cm}}
name & type & description \\
\hline
\texttt{n} & positive integer &   number of points for the angular grid in one sector; default: $20$\\\hline
\texttt{M} & positive integer &  number of points for the radial grid; default: $25$ \\\hline
\texttt{mu0\_2} & positive double & initial scale in GeV$^2$; default: $1$ GeV$^2$\\\hline
\texttt{muF\_2} & positive double & final scale in GeV$^2$; default: $10$ GeV$^2$\\\hline
\texttt{nstep} & positive integer & number of Runge-Kutta steps; default: $100$ \\\hline
\texttt{n\_thread} & positive integer & number of threads for kernel initialization; default: $1$ \\\hline
\texttt{charm\_THR} & positive double & charm threshold in GeV$^2$; default: $(1.27)^2 = 1.6129$ GeV$^2$\\\hline
\texttt{bottom\_THR} & positive double & bottom threshold in GeV$^2$; default: $(4.18)^2 = 17.4724$ GeV$^2$\\\hline
\texttt{xmin} & double $\in(0,1)$ & $\text{min}\lVert \bm{x} \rVert_{\infty}$; default: $10^{-2}$\\\hline
\texttt{interpolant\_type} & string: IT\_BOTH, IT\_PLUS, IT\_MINUS & whether to use the average of $\mathcal{I}^\pm$ or just $\mathcal{I}^+$ or $\mathcal{I}^-$ for the interpolation; default: IT\_BOTH\\\hline
\texttt{radial\_grid\_type} & string: \{LOG, PWR, IML, HYP\}\_GRID & version  of the radial grid; default: HYP\_GRID\\\hline
\texttt{angular\_grid\_type} & string: \{LIN, COS\}\_GRID & which version of the angular grid to use; default: LIN\_GRID\\\hline
\texttt{grid\_exponent} & positive double & exponent for the HYP and PWR grids; default: $3$\\\hline
\texttt{chiral\_even\_evolution} & boolean & whether to evolve the chiral-even sector; default: true \\\hline
\texttt{chiral\_odd\_evolution} & boolean & whether to evolve the chiral-odd sector; default: true \\\hline
\texttt{basefolder} & string & root folder in which the `dat' directory is searched;  both relative and absolute paths are accepted; default: "./"\\\hline
\texttt{kernel\_subfolder} & string & subfolder of `dat/kernels/' where the kernels are searched; default: "" \\\hline
\texttt{result\_subfolder} & string & subfolder of `dat/results/' where the results are to be stored; default: "" \\\hline
\texttt{integration\_rule} & string: GK\_\{21, 31, 41, 61\} & which Gauss-Kronrod rule to use; default: GK\_21\\\hline
\texttt{model} & string & unused; default: ""\\
\end{longtable}
\end{center}

An empty option raises a warning and the corresponding value is fixed to the default.
Each value \textbf{cannot contain}: white spaces, commas, colons, semi-colons, tabs or equal sign. If it does, a warning is printed that informs how the characters beyond the special character have been parsed into tokens. If the warning is printed, please check the configuration file. Moreover, \# can be used for comments.
A general valid entry in the configuration file (with * any number of the special characters above) is \texttt{*n*10*\#}.

\section{Configuration file for FORTRAN implementation}
\label{app_config_FORTRAN}
The initialization parameters for the FORTRAN code are kept in the text files, which have a numbered lines. They are summarized in the following table
\begin{center}
\begin{tabular}{c|c|p{10cm}}
\# & type & Description 
\\\hline
0  & logical & Flag to print information on the initialization process, and internal parameters.
\\\hline
1 & integer & Number of nodes in the angular sector, i.e. $n$ in (\ref{grid1})
\\\hline
2 & integer & Number of nodes in the along the radius, i.e. $M$ in (\ref{grid1})
\\\hline
3 & real & Minimal radius included in the grid
\\\hline
4 & real & Tolerance parameter used for zero estimation
\\\hline
5 & real & Tolerance for integration
\\\hline
6 & real & Maximal step of Runge-Kutta procedure (in $\ln(\mu)$)
\\\hline
7 & integer & Maximal number of processors allowed to use
\\\hline
8 & logical & Flag indicating that Chiral-Even kernels are to be used
\\\hline
9 & logical & Flag indicating that Chiral-Odd kernels are to be used
\\\hline
10 & logical & Flag indicating that singlet-part of evolution accounted (i.e. mixing between flavors)
\\\hline
11 & real & Mass of charm threshold.
\\\hline
12 & real & Mass of bottom threshold.
\\\hline
13 & logical & Flag indicating that kernels are to be read from files.
\\\hline
14 & string & Path to directory with kernels.
\end{tabular}
\end{center}

\bibliographystyle{JHEP}
\providecommand{\href}[2]{#2}\begingroup\raggedright\endgroup

\end{document}